\documentclass[12pt]{article}

\usepackage{epsf, amssymb}
\usepackage{cite}
\usepackage{epsfig}
\usepackage{amsmath}
\usepackage{latexsym}
\usepackage{epsfig}
\usepackage[english]{babel}
\usepackage{graphicx,color}
\usepackage{url}

\usepackage[colorlinks=true, linkcolor=blue, bookmarks=true]{hyperref}
\newcommand{\arXiv}[1]{\href{http://www.arXiv.org/abs/#1}{#1}}

\makeatletter
\renewcommand\section{\@startsection {section}{1}{\z@}%
                               {-3.5ex \@plus -1ex \@minus -.2ex}
                               {2.3ex \@plus.2ex}%
                               {\normalfont\large\bfseries}}
\renewcommand\subsection{\@startsection{subsection}{2}{\z@}%
                                 {-3.25ex\@plus -1ex \@minus -.2ex}%
                                 {1.5ex \@plus .2ex}%
                                 {\normalfont\bfseries}}
\makeatother

\def\IZ{\relax\ifmmode\mathchoice
{\hbox{\cmss Z\kern-.4em Z}}{\hbox{\cmss Z\kern-.4em Z}}
{\lower.9pt\hbox{\cmsss Z\kern-.4em Z}} {\lower1.2pt\hbox{\cmsss
Z\kern-.4em Z}}\else{\cmss Z\kern-.4em Z}\fi}
\def\IR{\relax{\rm I\kern-.18em R}}

\def\one{{\hbox{ 1\kern-.8mm l}}}

\def\tr{{\rm tr\,}}

\newlength{\bredde}
\def\slash#1{\settowidth{\bredde}{$#1$}\ifmmode\,\raisebox{.15ex}{/}
\hspace*{-\bredde} #1\else$\,\raisebox{.15ex}{/}\hspace*{-\bredde}
#1$\fi}

\newsavebox{\zzzbar}
\sbox{\zzzbar}
{\setlength{\unitlength}{0.9em}
\begin{picture}(0.6,0.7)
\thinlines
\put(0,0){\line(1,0){0.6}}
\put(0,0.75){\line(1,0){0.575}}
\multiput(0,0)(0.0125,0.025){30}{\rule{0.3pt}{0.3pt}}
\multiput(0.2,0)(0.0125,0.025){30}{\rule{0.3pt}{0.3pt}}
\put(0,0.75){\line(0,-1){0.15}}
\put(0.015,0.75){\line(0,-1){0.1}}
\put(0.03,0.75){\line(0,-1){0.075}}
\put(0.045,0.75){\line(0,-1){0.05}}
\put(0.05,0.75){\line(0,-1){0.025}}
\put(0.6,0){\line(0,1){0.15}}
\put(0.585,0){\line(0,1){0.1}}
\put(0.57,0){\line(0,1){0.075}}
\put(0.555,0){\line(0,1){0.05}}
\put(0.55,0){\line(0,1){0.025}}
\end{picture}}

\newcommand{\ena}{\end{eqnarray}}
\newcommand{\eq}[1]{(\ref{#1})}
\newcommand{\fig}[1]{Figure~\ref{#1}}

\newcommand{\be}{\begin{equation}}
\newcommand{\ee}{\end{equation}}

\def\be{\begin{equation}}
\def\ee{\end{equation}}

\def\r{\rho}

\def\({\left (}
\def\){\right )}
\def\[{\left [}
\def\[{\right ]}

\def\tr{\mathrm{tr}}

\def\ba{\begin{eqnarray}}
\def\ea{\end{eqnarray}}

\def \r{{\bf r}}

\input amssym.def
\input amssym.tex

\newcommand{\bbibitem}[1]{\bibitem{#1}\marginpar{#1}}
\def\Bibitem#1{\bibitem{#1}%
  \smash{\hbox to0pt{\raise1ex\hbox{\tiny[#1]}\hss}}}

\def\Label#1{\label{#1}%
  \smash{\hbox to0pt{\raise1ex\hbox{\tiny[#1]}\hss}}}
\def\noLabels{\let\Label=\label}
\def\nobbibitem{\let\bbibitem=\bibitem}
 \def\noBibitem{\let\Bibitem=\bibitem}

\def\[{\left [}
\def\]{\right ]}
\def\({\left (}
\def\){\right )}

\def\r{\rho}

\def\r2{\sqrt{2}}

\def\Label#1{\label{#1}%
  \smash{\hbox to0pt{\raise1ex\hbox{\tiny[#1]}\hss}}}
\def\noLabels{\let\Label=\label}
\def\nobbibitem{\let\bbibitem=\bibitem}

\newcommand{\bea}{\begin{eqnarray}}
\newcommand{\eea}{\end{eqnarray}}

\newcommand{\beq} {\begin{equation}}
\newcommand{\eeq} {\end{equation}}
\newcommand{\beqa} {\begin{eqnarray}}
\newcommand{\eeqa} {\end{eqnarray}}

\newcommand{\beqn}{\begin{eqnarray}}
\newcommand{\eeqn}{\end{eqnarray}}

\begin{document}

\begin{titlepage}
\begin{flushright}
arXiv:1212.6066
\end{flushright}
\vfill
\begin{center}
{\Large \bf Thermalization of the spectral function in strongly coupled two dimensional conformal field theories}

\vskip 10mm

{\large V.~Balasubramanian$^{a,b}$, A.~Bernamonti$^{c,d}$, B.~Craps$^d$, V.~Ker\"anen$^{e}$,\\
\vspace{3mm}
 E.~Keski-Vakkuri$^{f,g}$, B.~M\"uller$^{h}$, L.~Thorlacius$^{e}$, J.~Vanhoof$^{d}$}

\vskip 7mm

$^a$ David Rittenhouse Laboratory, Univ. of Pennsylvania, 
 Philadelphia, PA 19104, USA \\
$^b$Laboratoire de Physique Th\'{e}orique, \'{E}cole Normale Sup\'{e}rieure, 75005 Paris, France  \\
$^c$ Instituut voor Theoretische Fysica, KU Leuven,\\ 
Celestijnenlaan 200D, B-3001 Leuven, Belgium.\\
$^d$ Theoretische Natuurkunde, Vrije Universiteit Brussel, and \\
\hspace*{0.15cm}  International Solvay Institutes, 
Pleinlaan 2, B-1050 Brussels, Belgium. \\
$^e$Nordita, KTH Royal Institute of Technology and Stockholm University\\
Roslagstullsbacken 23, SE-106 91 Stockholm, Sweden\\
University of Iceland, Science Institute, 
Dunhaga 3, IS-107 Reykjavik, Iceland\\
$^f$Department of Physics, P.O.Box 64, FIN-00014 University of Helsinki, Finland\\
$^g$Department of Physics and Astronomy, Uppsala University, SE-75108 Uppsala, Sweden\\
$^h$Department of Physics, Duke University, Durham, NC 27708-0305, USA\\

\vskip 3mm
\vskip 3mm
{\small\noindent  {\tt vijay@physics.upenn.edu, alice@itf.fys.kuleuven.be, Ben.Craps@vub.ac.be, vkeranen@nordita.org, esko.keski-vakkuri@helsinki.fi,
mueller@phy.duke.edu, larus@nordita.org, joris.vanhoof@vub.ac.be}}

\end{center}
\vfill

\begin{center}
{\bf ABSTRACT}
\vspace{3mm}
\end{center}
Using Wigner transforms of Green functions, we discuss non-equilibrium generalizations of spectral functions and occupation numbers. We develop  methods for computing time-dependent spectral functions in conformal field theories holographically dual to thin-shell AdS-Vaidya spacetimes.

\end{titlepage}

\tableofcontents


\section{Introduction}
\label{intro}

The problem of thermalization of a quantum system that is prepared in a highly excited state is ubiquitous in physics. In contrast to situations where the equilibrium state of a quantum system is slightly perturbed and its return to equilibrium can be described by linear response theory, there is no general approach to the problem of thermalization starting far from equilibrium. In some cases, e.~g.\ when the system is dilute and composed of weakly coupled quasiparticles, kinetic theory provides a general solution. However, many quantum systems of interest are either dense or strongly coupled, or both, and kinetic theory does not apply. 

A widely studied example is the strongly interacting matter produced in high energy nuclear collisions \cite{Jacak:2012zz}. When thermally equilibrated, this matter is known as a quark-gluon plasma, but the collision is thought to produce an initial state composed of densely packed, highly excited, but locally coherent gauge field domains, often referred to as a ``glasma'' \cite{Lappi:2006fp}. The thermalization of this initial state is an open problem of intense phenomenological interest \cite{Dusling}.

The late time evolution of the energy-momentum tensor of a thermalizing quantum system is described by the effective low-energy theory of energy and momentum, fluid dynamics.  Chesler and Yaffe \cite{Chesler:2010bi} and Heller et al. \cite{Heller:2011ju,Heller:2012je}, have shown that fluid dynamics becomes a good approximation rather quickly.  But this ``hydroization''  of a quantum system does not imply full or even approximate thermalization of the system, because the anisotropy of the energy-momentum tensor may remain quite large, and there is no assurance that the expectation values of other observables are near their thermal values.

Time evolution from near-equilibrium initial states  in a strongly coupled gauge theory with a holographic AdS/CFT dual has been extensively studied (see the review \cite{Hubeny:2010ry}).  More recently, focus has moved to thermalization from far off-equilibrium initial states.  One interesting 
quantity to consider in such studies is the holographic entanglement entropy \cite{Ryu:2006bv}, which was generalized to time-dependent backgrounds in \cite{Hubeny:2007xt}, in
particular the collapse of a homogeneous massless shell in the AdS space-time, leading to black hole formation in the bulk, which corresponds to thermalization in the dual field theory \cite{Banks:1998dd}.
Thermalization was studied in more detail in \cite{AbajoArrastia:2010yt,Albash:2010mv,Balasubramanian:2010ce,Balasubramanian:2011ur, mutualinfo} 
following the time evolution of entanglement entropy, Wilson loops and equal-time correlation functions.   The most nonlocal quantity, the entanglement entropy, reached thermal equilibrium last, but still very fast, on a time scale saturating the causality bound \cite{Calabrese:2005in}.
A complementary approach to  thermalization was investigated in \cite{CaronHuot:2011dr,Chesler:2011ds,Chesler:2012zk}, who considered thermalization after an anisotropic deformation of the boundary
spacetime \cite{Chesler:2009cy}, in particular how fluctuations created near the horizon and dissipation  come to a balance  to satisfy a generalized fluctuation-dissipation theorem. 
Isotropization and thermalization were also investigated in \cite{Heller:2012km}, by considering a large set of bulk initial conditions. Rotationally symmetric isotropization including radial flow was recently investigated in \cite{vanderSchee:2012qj}.

Here we extend this previous work by considering time-like correlation functions. This allows us to generalize concepts usually studied in equilibrium to non-equilibrium states, and to compute them holographically. Two important quantities of this kind are the spectral density and the occupation number distribution of  states. Time-dependent versions of these quantities are discussed and illustrated in a simple example in section~\ref{spectral}. .

Our holographic computations will be carried out in the thin-shell AdS$_3$ Vaidya  space-time,
\begin{equation} \label{eq:Vaidya}
ds^{2}=- \left[r^{2}-\theta(v)R^{2}\right]dv^{2}+2dvdr+r^{2}dx^{2}.
\end{equation}
which models time evolution following the sudden injection of energy in the dual field theory.
We will develop two approaches for computing two-point functions with timelike separations.  The first approach, described in section~\ref{geodesic}, uses the geodesic approximation, which has been employed for computing two-point functions of spacelike separated heavy operators in \cite{AbajoArrastia:2010yt, Balasubramanian:2010ce,Balasubramanian:2011ur, Aparicio:2011zy}.  There are  significant new challenges in the present case --  real timelike geodesics do not reach the boundary of AdS space (and therefore cannot be used to connect timelike separated boundary points).   An alternative approach could have been to use analytic continuation from Euclidean signature, but  AdS-Vaidya does not allow a standard continuation of this kind.  Indeed, the absence of a conventional Euclidean continuation with which to define Green functions is a general challenge in the study of non-equilibrium systems.

Here we propose  novel approaches to overcoming these obstacles within the AdS/CFT correspondence:  (i) by defining a new, non-standard Euclidean continuation, (ii) by using complex geodesics,  (iii) by developing a prescription for splicing propagators across the infalling shell.    In a saddle point limit appropriate to heavy operators, we obtain an analytic expression for equal-space correlators that agrees precisely between all three approaches.    Method (iii) does not require the probe to be heavy, and hence in 
 Sec.~\ref{beyond} we use it to  obtain numerical results for retarded two-point functions in momentum space.   

A complementary approach to the holographic non-equilibrium spectral function and the fluctuation-dissipation relation has been developed in \cite{Banerjee:2012uq,Mukhopadhyay:2012hv}. They consider perturbative expansions (derivative and amplitude)  away from thermal equilibrium, working first at leading order \cite{Banerjee:2012uq} (where they also study non-equilibrium
shift effects)  and then extending to higher orders \cite{Mukhopadhyay:2012hv}. In particular \cite{Mukhopadhyay:2012hv} also derives a  generalized (holographic) fluctuation-dissipation relation for Wigner transformed spectral function and statistical function (the anticommutator Green function) which holds for non-equilibrium states which are perturbatively connected
to thermal equilibrium.  It would be interesting and important to compare the approach of \cite{Banerjee:2012uq,Mukhopadhyay:2012hv}  to the methods and context of this paper.

\setcounter{equation}{0}

\section{Time-dependent spectral functions and occupation numbers}
\label{spectral}


\subsection{Spectral function and occupation numbers in equilibrium}
\label{equilibrium}

Consider a system prepared in a state described by a density matrix $\hat\rho $. In the interaction picture or in the Heisenberg picture, the density matrix is
time-independent while the operators  evolve in time.
In order to study how thermalization proceeds, we will look at the evolution of  correlation functions.  The relevant correlators are the time-ordered (Feynman) correlation
funct
\be i G_{\rm F}({\bf x},t) = \tr \langle \hat\rho\, T\big({\cal O}({\bf x},t) {\cal O}(0,0)\big) \rangle \,,
\ee
and the retarded correlation function
\be i G_{\rm R}({\bf x},t) = \theta(t) \tr \langle \hat\rho\, [{\cal O}({\bf x},t), {\cal O}(0,0)] \rangle\,.
\ee
If the system is translation invariant in space and time,
we can consider the Fourier transforms of the correlation functions $G_{\rm
F}({\bf k},\omega)$ and $G_{\rm R}({\bf k},\omega)$.
In Fourier space, the retarded correlation function is complex valued, and its imaginary part defines the
spectral function\footnote{Thermal spectral functions were first studied in the AdS/CFT correspondence for ${\cal N}=4$ super-Yang-Mills theory in \cite{Kovtun:2006pf,Teaney:2006nc}. }
\be \label{eq:defrho}
\rho({\bf k},\omega) \equiv
-2\, {\rm Im}\, G_{\rm R}({\bf k},\omega) .
\ee
The imaginary part of the time-ordered correlator contains additional information, which depends on the mean occupation
number density $n({\bf k},\omega)$ of the Fourier modes.
When the system is at thermal equilibrium, the correlation functions are related by the Fluctuation-Dissipation
Theorem (FDT)
\be \label{eq:defn}
[1+ 2 n({\bf k},\omega)]\, \rho({\bf k},\omega) = - 2\, {\rm Im}\, G_{\rm F}({\bf k},\omega) \ ,
\ee
with the occupation number equal to the Bose (or
Fermi) distribution $n({\bf k},\omega)=(e^{\omega/T} \mp 1)^{-1}$.   The authors of  \cite{CaronHuot:2011dr} have emphasized the usefulness of (\ref{eq:defn}) as a measure of the degree of thermalization in holographic theories,  in the context of a model where fluctuations are created near the stretched horizon and transported to the
boundary. In these works the fluctuation-dissipation theorem has also been expressed as a relation between the symmetrized and antisymmetrized correlation functions
\be
 G_{\rm S} ({\bf x},t) = \tr
\langle \hat\rho\, \{{\cal O}({\bf x},t), {\cal O}(0,0)\} \rangle \,, \qquad
G_{\rm A} ({\bf x},t) = \tr \langle \hat\rho\, [{\cal O}({\bf x},t), {\cal O}(0,0)] \rangle \,,
 \ee
so that
\be G_{\rm S}({\bf k},\omega) =  [1+ 2 n({\bf k},\omega)]\, G_{\rm A}({\bf k},\omega) \ .
\ee


\subsection{Time-dependent generalizations} \label{time}

In order to determine the degree of thermalization we would like to define time-dependent generalizations
 of the occupation number density $ n({\bf k},\omega)$ and the spectral function $\rho({\bf k},\omega)$.  By definition, a time-dependent system in not  time translation invariant, so the Feynman and retarded correlators explicitly depend on two instants of time:
$ G_{\rm F,R}({\bf x}_2,t_2;{\bf x}_1,t_1)$. Space is still homogenous, so the correlators depend only on ${\bf x}_1-{\bf x}_2$ which can be Fourier transformed to momentum ${\bf k}$.  But a simple transform from time to frequency space $\omega$ is no longer meaningful as we seek a time-{\it dependent} notion of occupation number and spectral density.  We therefore resort to a well-known alternative  -- the Wigner transform.

First introduce an average time $T$ and a relative time $t$ by
\be t_1 \equiv T-\frac{t}2,\ \ \ \ \ t_2 \equiv T+\frac{t}2, \ee
and define $T$-dependent Green functions by
keeping the average time $T$ fixed and Fourier transforming in the relative time $t$:
\be \label{eq:Wigner}
G_{\rm R,F}(T,\omega)\equiv\int_{-\infty}^\infty dt \, e^{i\omega t}\,G_{\rm R,F}(T,t) \,,
\ee
where we have suppressed the spatial coordinates. The $T$-dependent spectral function is then defined as
\be \label{eq:rhoT}
\rho(T,\omega)\equiv -2\,{\rm Im}\,  G_{\rm R}(T,\omega) \,.
\ee

In this paper, our goal is to study how the spectral function evolves as a function of the average time $T$. In order
to focus on information near $T$, we may consider small intervals of relative time $t$ around $t=0$. To this end, we add a Gaussian window function into the integrand\footnote{This is similar in spirit to the Husimi distribution, which is a Gaussian smearing of the Wigner distribution on phase space \cite{WignerHusimi}.
 It is also related to a well known method in signal analysis. For a signal $s(t)$ with time dependent frequency,
it is convenient to add a Gaussian weight factor into the integrand with a small window about a mean time instant $t'$,
\be\label{eq:ssigma} s_\sigma(t',\omega) \equiv \int_{-\infty}^\infty dt \, e^{i\omega t}\,e^{-\frac{(t-t')^2}{2\sigma^2}}\,s(t) \ .
\ee
Then by studying $s_{\sigma}(t',\omega)$ as a function of $t'$ one can track changes in frequency. This procedure is known as the Gabor transform,  and it is a special case of a more general class
of wavelet transformations (with more customized window functions called wavelets).
In our case, we are studying functions of two time variables, and we are primarily interested in tracking changes
as the untransformed variable $T$ changes, hence we are setting $t'=0$. It would be an additional possibility to perform a full
Gabor transform, in which case varying $t'$ would allow us to study information at different time separations, but that is beyond
the scope of this paper.}:
\be\label{eq:Gsigma} G_\sigma(T,\omega) \equiv \int_{-\infty}^\infty dt \, e^{i\omega t}\,e^{-\frac{t^2}{2\sigma^2}}\,G(T,t) \, ,
\ee
where $\sigma$ controls the width of the time window,
and $G$ is a correlator.   In the presence of a finite time window the $T$-dependent spectral function becomes
\be \label{eq:rhosigma}
  \rho_{\sigma} (T,\omega ) \equiv -2\, {\rm Im}\, G_{\sigma,\rm R}(T,\omega ) \ .
\ee

We can now define a generalized, time-dependent notion of occupation number density by the relation
\be \label{eq:nsigma}
 n_{\sigma}({\bf k},T,\omega)  \equiv \frac{1}{2} \left[ \frac{{\rm Im}\,
 G_{\sigma, \rm F}({\bf k},T,\omega)}{{\rm Im}\, G_{\sigma, \rm R}({\bf k},T,\omega)} -1 \right]  \, .
\ee
In an interacting field theory we expect the occupation number density of the non-equilibrium system to evolve towards thermality  $(e^{\beta \omega}-1)^{-1}$ (for bosons), so that the system eventually satisfies the fluctuation-dissipation theorem and becomes fully thermalized.

We note that the notion of occupation number density is only meaningful for momenta and frequencies where the spectral function has support: if a mode does not exist in the spectrum, it is meaningless to ask whether it is occupied. In physical quantities, the occupation number density will always be accompanied by the spectral function, so that no contributions originate from regions in momentum space where the spectral function vanishes. This is reflected in \eq{eq:defn}. The issue does not arise in the conformal field theories considered later in the paper but the simple quantum mechanical example in the following section has a discrete spectrum and in that case expressions like \eq{eq:nsigma} need to be interpreted with care.

As explained in appendix~\ref{app:entropy}, we can now follow the analysis in \cite{BPM} to define a time-dependent entropy density
\be
S(T) = \int d\omega \, d{\bf k} \, \rho({\bf k}, T,\omega) \, s({\bf k}, T, \omega)
\ee
where
\be
s({\bf k}, \omega, T) =
\left( n({\bf k},T, \omega) + 1\right)
\ln \left( n({\bf k},T, \omega) + 1\right)
-
\left( n({\bf k},T, \omega)  \right)
\ln \left( n({\bf k},T, \omega)\right).
\ee
This definition is a natural extension of the conventional microcanonical definition of von Neumann entropy of a system, and reduces  to the latter in equilibrium.  It should be contrasted with  more specialized notions of entropy such as entanglement and Renyi entropies, which have been the subject of many studies of holographic thermalization.


\subsection{A simple example: the quenched harmonic oscillator} \label{harmonic}

To illustrate the out-of-equilibrium notion of a time-dependent spectral function in \eqref{eq:rhoT}, and in \eqref{eq:rhosigma} when a finite time window is introduced, we first consider the harmonic oscillator with characteristic frequency $\omega_0$. 

As we will eventually be interested in applying our methods to quantum field theories, it is useful to
think about the harmonic oscillator as a $0+1$ dimensional quantum field theory. This quantum field theory has a single one particle state $a^{\dagger}|0\rangle$ with energy $\omega_0$. Multiparticle states then appear as the excited states $(a^{\dagger})^n|0\rangle$. We will study the single particle spectral function obtained from the two point function of $x(t)$ and the occupation number of the single energy eigenstate. It should be also noted that all of our results in this section generalize trivially to free quantum field theories with the substitution
\beq
\omega_0\rightarrow \sqrt{{\bf k}^2+m^2}.
\eeq
This is because free field theories are just a set of decoupled harmonic oscillators labeled by a spatial momentum ${\bf k}$, and because the observables we consider are related to the two 
point correlation functions, which are diagonal in $k$-space.

The retarded Green function for the oscillator in its ground state is
\be\label{retarded}
G_R(t)=-\frac{\sin(\omega_0t)}{\omega_0}\,\theta(t).
\ee
Writing
\be\label{heaviside}
\theta(t)=\frac{i}{2\pi}\int_{-\infty}^\infty d\omega \frac{e^{-i\omega t}}{\omega+i0^+},
\ee
we find the Fourier transform
\be
G_R(\omega)\equiv\int_{-\infty}^\infty dt \, e^{i\omega t}\,G_R(t)=\frac{1}{2\omega_0}\left(\frac{1}{\omega-\omega_0+i0^+}-\frac{1}{\omega+\omega_0+i0^+}\right).
\ee
Using
\be\label{principal}
\frac{1}{\omega+i0^+}=P\left(\frac1\omega\right)-i\pi\delta(\omega),
\ee
the spectral density is given by
\be
\rho(\omega)\equiv -2 \,{\rm Im}\, G_R(\omega)=\frac{\pi}{\omega_0}\,\left[\delta(\omega-\omega_0)-\delta(\omega+\omega_0)\right].
\ee


Following the general discussion of Section~\ref{time}, we now introduce a time window into the Green function,
\be\label{eq:Gsigma2}
G_{\sigma, R} (\omega)\equiv \int_{-\infty}^\infty dt \, e^{i\omega t}\,e^{-\frac{t^2}{2\sigma^2}}\,G_R(t) \ .
\ee
Using \eq{retarded}, \eq{heaviside} and the known Fourier transform of a Gaussian, we find
\be
G_{\sigma,R} (\omega)= \int_{-\infty}^\infty \frac{d\omega'}{2\omega_0}\,\frac1{\omega'{+}i0^+}\,\frac{\sigma}{\sqrt{2\pi}}\,
\left[ \exp\left(-\frac{\sigma^2}2(\omega'{-}\omega{+}\omega_0)^2 \right) - \exp\left(-\frac{\sigma^2}2(\omega'{-}\omega{-}\omega_0)^2 \right)\right].
\ee
The spectral function is then
\be
\rho_\sigma(\omega)\equiv -2 \,{\rm Im}\, G_{\sigma,R}(\omega) \ ,
\ee
and using \eq{principal}, we find
\bea
\rho_\sigma(\omega)&=&\frac{\pi}{\omega_0}\,\left[\frac{\sigma}{ \sqrt{2\pi}} \exp\left(-\frac{\sigma^2}2(\omega
-\omega_0)^2\right)-\frac{\sigma}{ \sqrt{2\pi}} \exp\left(-\frac{\sigma^2}2(\omega+\omega_0)^2\right)\right]\cr
&\equiv&
\frac{\pi}{\omega_0}\,\left[\delta_{\sigma}(\omega-\omega_0)-\delta_{\sigma}(\omega+\omega_0)\right].
\eea
The Gaussian time window simply results in a Gaussian smearing of the delta function peaks in the spectral density.


Next we study a quenched harmonic oscillator, with frequency $\omega_i$ for $t<t_0$ and frequency $\omega_f$ for $t>t_0$. The time of the quench can be taken to be $t_0=0$ without loss of generality.
The retarded Green function $G_R(t_2,t_1)$ now depends on both $t_1$ and $t_2$, not only on their difference. It is given by
\bea \nonumber
G_R(t_2,t_1)=& \left\{\begin{array}{ccl}
-\frac1{\omega_i}\sin[\omega_i(t_2-t_1)]\,\theta(t_2-t_1) & \textrm{if} & t_1,t_2<0 \,, \\
-\frac1{\omega_f}\sin[\omega_f(t_2-t_1)]\,\theta(t_2-t_1) & \textrm{if} & t_1,t_2>0 \,, \\
\frac1{\omega_i}\sin[\omega_it_1]\cos[\omega_ft_2]-\frac1{\omega_f}\cos[\omega_it_1]\sin[\omega_ft_2]  & \textrm{if} &  t_1<0<t_2\,, \\
0 & \textrm{if} & t_2<0<t_1 \,.
\end{array}\right.
\\
\eea
The matching conditions at $t_0=0$ are such that for $t_1<0$ both $G_R$ and $\partial G_R / \partial t_2$ are continuous across $t_2=0$, and for $t_2>0$ both $G_R$ and $\partial G_R / \partial t_1$ are continuous across $t_1=0$.

In terms of the average time and relative time coordinates, $t_1 \equiv T-t/2,\ t_2  \equiv T+t/2$, the retarded Green function reads
\bea\label{GTt}
G_R(T,t)= \left\{\begin{array}{lcl} \nonumber
-\frac1{\omega_i}\sin(\omega_it)\,\theta(t) & \textrm{if} & T<0,\> \frac t2<- T \\
-\frac1{\omega_f}\sin(\omega_ft)\,\theta(t) & \textrm{if} & T>0,\> \frac t2<T \\
-\frac{1}{\omega_i}\sin[\omega_i(\frac t2{-}T)]\cos[\omega_f(\frac t2{+}T)] \theta(t) & \   & \  \\
 \quad -\frac{1}{\omega_f}\sin[\omega_f(\frac t2{+}T)]\cos[\omega_i(\frac t2{-}T)] \theta(t)& \textrm{if} & -\frac t2<T<\frac t2
\end{array}\right.
\\
\eea
We can then study its Wigner transform \eqref{eq:Wigner} and obtain the time-dependent spectral function \eqref{eq:rhoT}. The main features of $\rho(T,\omega)$ can be uncovered by inspecting our expression \eq{GTt} for $G(T,t)$.
Consider, for instance, the very early time regime, $T\rightarrow -\infty$. In this limit, $G(T,t)$ is given by the first line in \eq{GTt}, and $\rho(T,\omega)$ reduces to the spectral function for a time-independent harmonic oscillator with frequency $\omega_i$. Similarly, in the late time limit $T\rightarrow\infty$, we recover the spectral function of a harmonic oscillator with frequency $\omega_f$. The time-dependent spectral function for a quenched harmonic oscillator thus interpolates between those of harmonic oscillators with the initial and final frequencies. To get a first idea of the intermediate-time regime, we can take the average time to be the transition time, $T=t_0=0$. In this case \eq{GTt} reduces to
\be
G_R(0,t)=-\frac12\left[\left(\frac1{\omega_i}+\frac1{\omega_f}\right)\sin[(\frac{\omega_i+\omega_f}{2})t]
+\left(\frac1{\omega_i}-\frac1{\omega_f}\right)\sin[(\frac{\omega_i-\omega_f}{2})t]\right]\theta(t) \,.
\ee
and we see that at $T=0$ the retarded Green function executes beats between the initial and final frequencies in relative time.

The Fourier transform of the retarded Green function with respect to relative time can be carried out explicitly for general values of the average time. Taking the imaginary part then gives the following time-dependent spectral function,
\bea\label{eq:rhoTquenched}
\rho(T,\omega) &= \left\{\begin{array}{lcl}
\frac{\omega_i^2-\omega_f^2}{\omega_i}
\left(\frac{\sin{\left[2T(\omega-\omega_i)\right]}}{(\omega-\omega_i)((2\omega-\omega_i)^2-\omega_f^2)}
- \frac{\sin{\left[2T(\omega+\omega_i)\right]}}{(\omega+\omega_i)((2\omega+\omega_i)^2-\omega_f^2)}\right)
\>& \textrm{if} & T<0 \\
\pi \left(\frac1{\omega_i}+\frac1{\omega_f}\right)\bigl(\delta{(2\omega-\omega_i+\omega_f)}-\delta{(2\omega+\omega_i+\omega_f)}\bigr) & \ & \ \\
+ \pi \left(\frac1{\omega_i}-\frac1{\omega_f}\right)\bigl(\delta{(2\omega-\omega_i-\omega_f)}-\delta{(2\omega+\omega_i-\omega_f)}\bigr)  \>& \textrm{if} & T=0 \\
\frac{\omega_i^2-\omega_f^2}{\omega_f}
\left(\frac{\sin{\left[2T(\omega-\omega_f)\right]}}{(\omega-\omega_f)((2\omega-\omega_f)^2-\omega_i^2)}
- \frac{\sin{\left[2T(\omega+\omega_f)\right]}}{(\omega+\omega_f)((2\omega+\omega_f)^2-\omega_i^2)}\right)
\>& \textrm{if} & T>0
\end{array}\right.
\eea
We observe that the time-dependent definition \eqref{eq:rhoT} allows for negative values of the spectral function for $\omega >0$.
This is to be compared with the equilibrium spectral function that is positive definite for positive $\omega$.

\begin{figure}[h]
\begin{center}
\includegraphics[width=0.45 \textwidth]{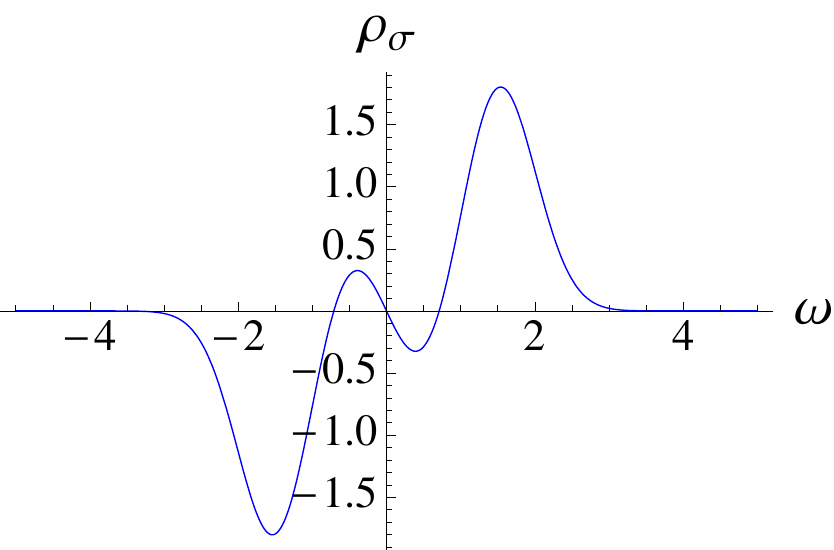}
\hfil
\includegraphics[width=0.45 \textwidth]{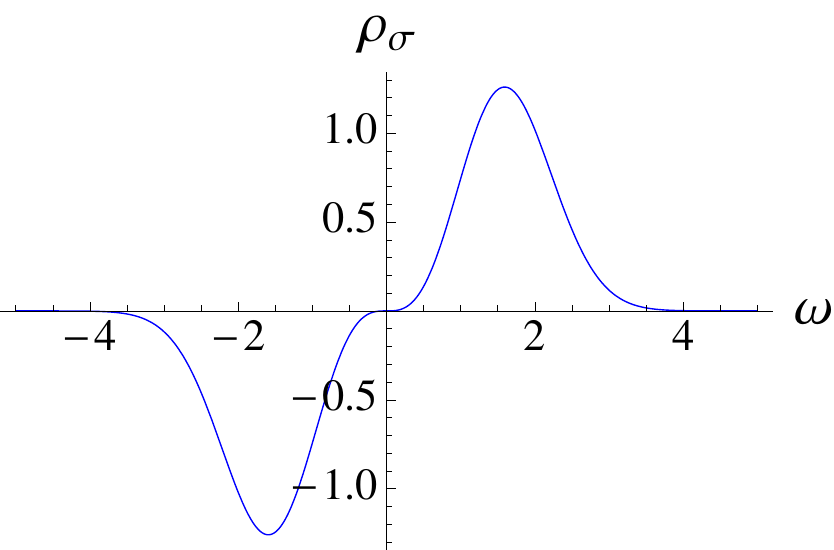}
\caption{Spectral function of a quenched harmonic oscillator ($\omega_i=2$, $\omega_f=1$) with Gaussian time window $\rho_\sigma(T, \omega)$ at $T=0$ for $\sigma =2$ (left) and $\sigma \approx 1.48$ (right).
}
\label{fig:rhosigma}
\end{center}
\end{figure}

Although it is not necessary, it may be useful to introduce time windows as in \eqref{eq:Gsigma} in order to define generalized time-dependent spectral functions that only depend on the behavior near $T$. This will have a smearing effect, for instance on the various delta functions appearing in our quenched harmonic oscillator example as can be seen in Fig.~\ref{fig:rhosigma}. Observe in particular that the spectral function can be made positive definite for $\omega \ge 0$ for an appropriate value of the width $\sigma$ (Fig.~\ref{fig:rhosigma} right). This is similar to the situation with conventional phase space Wigner density which attempts to define a notion of particle density in the phase space of a quantum theory.   The Wigner density is sometimes negative, but the associate Husimi density, obtained by smearing with an appropriate Gaussian is everywhere positive \cite{WignerHusimi}.

\begin{figure}[h]
\begin{center}
\begin{tabular}{ccc}
\includegraphics[width=0.3 \textwidth]{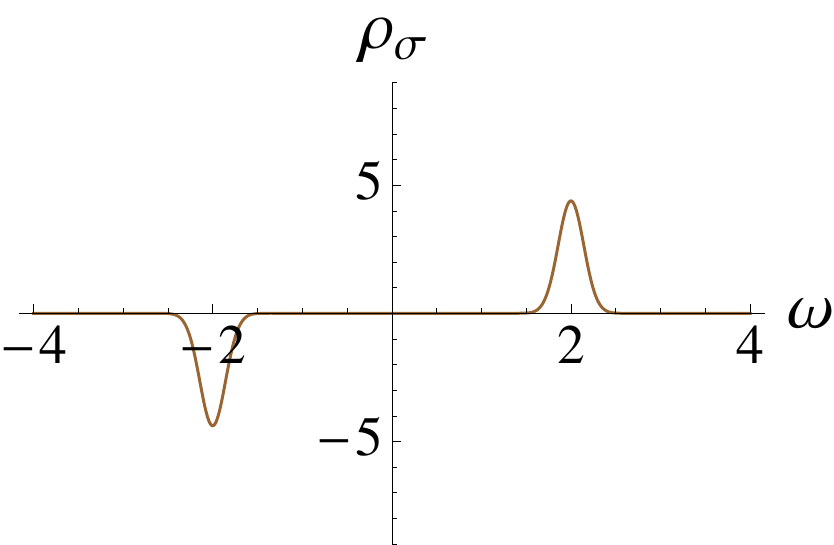} & \includegraphics[width=0.3 \textwidth]{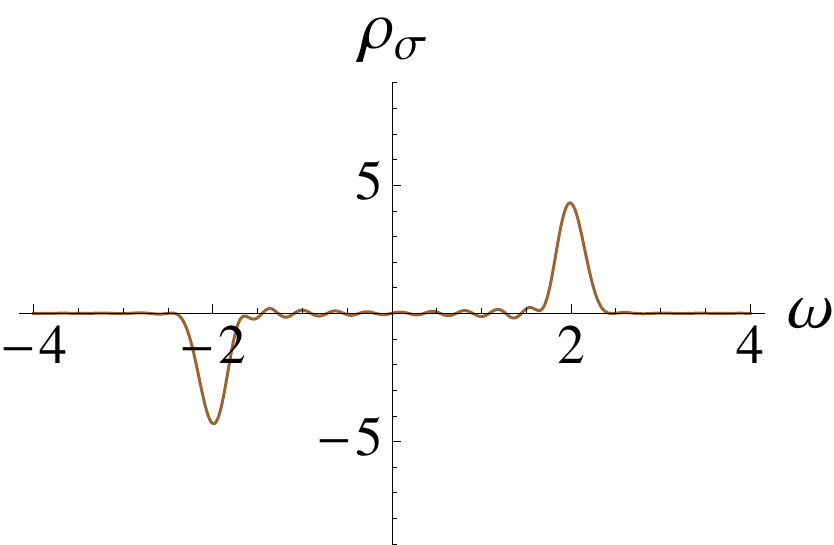}  & \includegraphics[width=0.3 \textwidth]{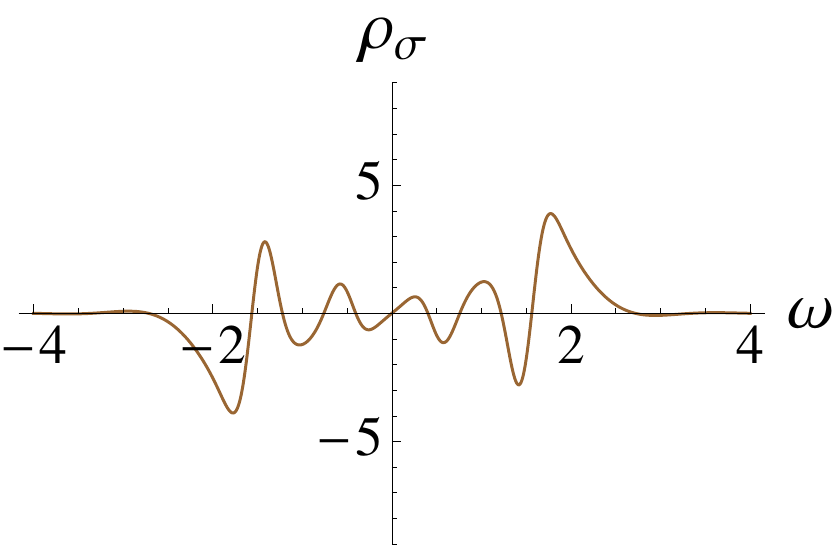}  \\
\includegraphics[width=0.3 \textwidth]{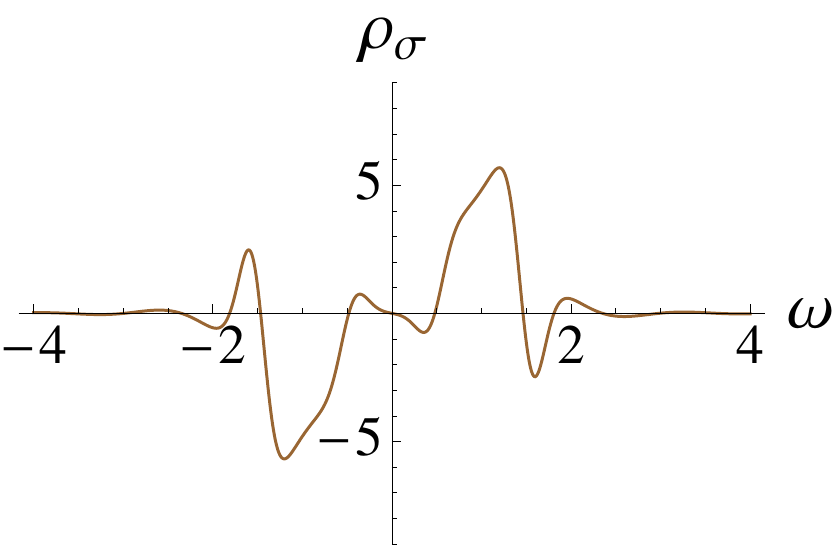} & \includegraphics[width=0.3 \textwidth]{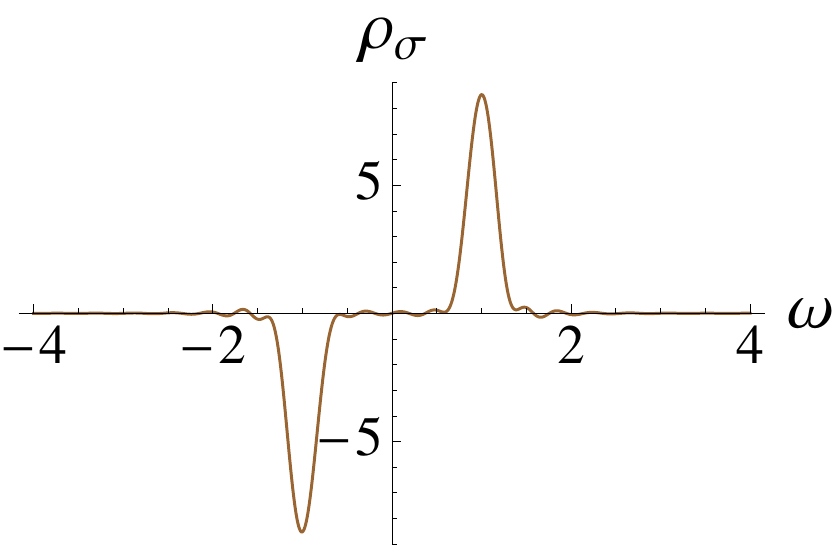} &\includegraphics[width=0.3 \textwidth]{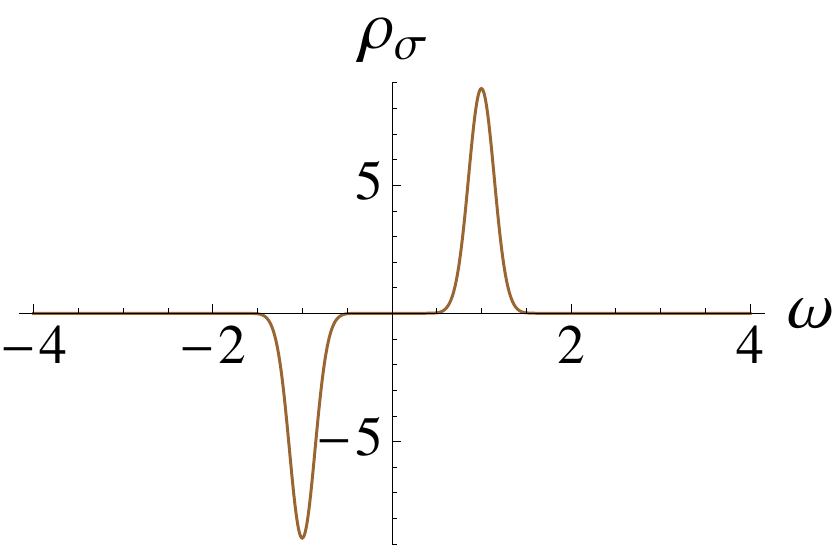} \\
\end{tabular}
\caption{Time evolution of $\rho_\sigma(T, \omega)$ (with $\omega_i=2$, $\omega_f=1$) for $\sigma = 7$ and $T = -12, -7.2, -2.4, 2.4, 7.2, 12$ from left to right. At early (late) times the spectral function is the one of a time-independent harmonic oscillator with frequency $\omega_i$ ($\omega_f$).
}
\label{fig:rhoT}
\end{center}
\end{figure}

Examining the smeared spectral function for different times (Fig.~\ref{fig:rhoT}) shows that $\rho_\sigma$ evolves between the spectral functions of harmonic oscillators with the correct initial and final frequencies.

\begin{figure}[h]
\begin{center}
\includegraphics[width=0.45 \textwidth]{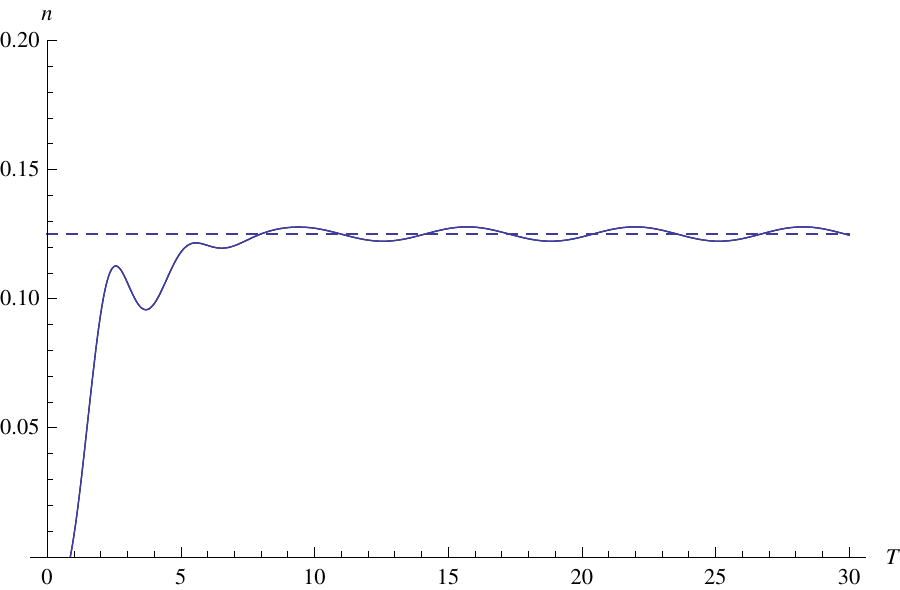}
\\
\caption{The plot shows the occupation number for frequency $\omega = \omega_f = 1$ as a function of average time (here $T$ is multiplied
by a factor of 2 compared with the convention in the text) for a quenched harmonic oscillator, with final frequency $\omega_f = 1$ and initial frequency
$\omega_i = 2$, with smearing parameter $\sigma = \sqrt{8}$.   The limiting values agrees with a conventional computation of the final occupation number in terms of a Bogoliubov transformation of the initial vacuum.
}
 \label{fig:occupation}
\end{center}
\end{figure}

Finally, we can compute a time-dependent occupation number, as defined in \eq{eq:nsigma}. The result for specific parameter values is shown in \fig{fig:occupation}. The occupation number quickly evolves to its asymptotic value, which is given by the expectation value of the harmonic oscillator number operator (for oscillator frequency
$\omega_f$) evaluated in the ground state of an oscillator of frequency $\omega_i$, which is the state of the system at the time of the quench. The oscillations around the final value arise from the smearing due to the finite time window - their size depending on the choice of the parameter $\sigma$.

\setcounter{equation}{0}

\section{The geodesic approximation in the AdS/CFT correspondence}\label{geodesic}

To use the methods developed above to discuss  time dependent spectral functions in the AdS/CFT correspondence we need to first compute two point functions in the latter setting.     In this section, we use a geodesic approximation to  compute two-point functions with timelike separation in a two dimensional CFT dual to the AdS$_3$-Vaidya spacetime \eqref{eq:Vaidya}.   This approximation is accurate for heavy operators in the CFT.   We will find simple analytic results for the special case of equal-space two-point functions, but our methods generalize to generic separations. These results allow us to compute an integral over spatial momenta of the time-dependent spectral function of high-dimension operators, as well as a weighted integral of time-dependent occupation numbers.

Consider a scalar operator $\mathcal{O}(x,t)$ with conformal dimension $\Delta$ in the dual CFT. The time-ordered two-point function $\langle\mathcal{O}(x_{1},t_{1})\mathcal{O}(x_{2},t_{2})\rangle$ is given by a path integral over all paths $\mathcal{P}$ that connect the two insertion points $(x_{1},t_{1})$ and $(x_{2},t_{2})$ on the boundary:
\begin{equation}
\langle\mathcal{O}(x_{1},t_{1})\mathcal{O}(x_{2},t_{2})\rangle=\int\mathcal{D}\mathcal{P} \, e^{-\Delta L(\mathcal{P})}
\qquad\text{with}\qquad
L(\mathcal{P}) \equiv \int_{\mathcal{P}}\sqrt{g_{\mu\nu}\frac{dx^{\mu}}{d\lambda}\frac{dx^{\nu}}{d\lambda}}\text{d}\lambda\,.
\end{equation}
For large $\Delta$, we can use a saddle point approximation, in which the sum over all paths can be approximated by a sum over all geodesics connecting the boundary endpoints \cite{Balasubramanian:1999zv}
\begin{equation}
\langle\mathcal{O}(x_{1},t_{1})\mathcal{O}(x_{2},t_{2})\rangle\sim\sum_{\text{geodesics}}e^{-\Delta\mathcal{L}}\,,
\end{equation}
where ${\cal L}$ denotes the geodesic length.
However, due to contributions near the asymptotically AdS boundary, the geodesic length between two boundary points contains a universal divergence and needs to be renormalized. Throughout this Section, we will
define a renormalized length $\delta {\cal L} \equiv {\cal L} - 2 \ln (r_0)$, in terms of the bulk cut-off $r_0$, by removing the divergent part of the geodesic length in pure AdS (see Eq.~\eqref{eq:AdSrenorm}-\eqref{eq:AdSlength} in Appendix~\ref{app:geodesics}). The renormalized two point function can then be approximated by
\begin{equation}
\langle\mathcal{O}(x_{1},t_{1})\mathcal{O}(x_{2},t_{2})\rangle_{\text{ren}}\sim e^{-\Delta \delta\mathcal{L} }\,,
\end{equation}
where $\delta\mathcal{L}$ is the renormalised length of the geodesic that connects the points $(x_{1},t_{1})$ and $(x_{2},t_{2})$ on the boundary.

The geodesic approximation works most straightforwardly in Euclidean spacetimes, or in static spacetimes that can be Wick rotated to Euclidean signature. In \cite{Balasubramanian:2010ce,Balasubramanian:2011ur}, this approach was used in a time-dependent setting,  to determine equal-time two point functions in the field theory dual of the Vaidya spacetime. This work was further extended in \cite{Aparicio:2011zy} to non-equal-time two-point functions with spacelike separations. In all these cases, the relevant bulk geodesics $\mathcal{P}$ were spacelike, and the geodesic length $\mathcal{L}(\mathcal{P})$ was real.

New complications occur when the points on the boundary are timelike separated, in which case we might expect to connect them using timelike geodesics.  The main problem is that (real) timelike geodesics in an asymptotically AdS spacetime never extend to the boundary. For example, in the case of empty AdS$_{3}$ with metric
\begin{equation}
ds^{2}=-r^{2}dt^{2}+\frac{dr^{2}}{r^{2}} +r^{2}dx^{2}\,,
\end{equation}
the timelike geodesics ($e^{2}>j^{2}$):
\begin{equation}
\begin{cases}
r(\lambda)=\sqrt{e^{2}-j^{2}}\cos(\lambda) \\
x(\lambda)=x_{0}+\frac{j}{e^{2}-j^{2}}\tan(\lambda) \\
t(\lambda)=t_{0}+\frac{e}{e^{2}-j^{2}}\tan(\lambda)
\end{cases}
\end{equation}
satisfy $r(\lambda)\leq\sqrt{e^{2}-j^{2}}$, $\forall \lambda$.  Thus they do not reach the boundary, where $r \to \infty$.

In fact, this is a specific realization of the general difficulty of defining correlation functions in time-dependent spacetimes that do not have a  Euclidean continuation.     Given that a general prescription is not available, we will here propose and investigate three potential approaches:
\begin{enumerate}
\item For time-independent spacetimes the standard prescription for vacuum correlators is to calculate in Euclidean signature and then analytically continue to Lorentzian signature.  Since there is no obvious Wick rotation of the AdS-Vaidya spacetime, we will propose and use a non-standard Euclidean continuation in Sec.~\ref{sect:Euclidean}.
\item We will consider complexified geodesics in  Sec.~\ref{sect:complex}.  In this approach, geodesics between timelike separated boundary points make excursions in a complexified AdS-Vaidya geometry before returning to real points on the boundary.
\item In Sec.~\ref{sect:joining} we will directly connect the AdS  and black hole retarded bulk-to-boundary propagators across the infalling shell, and will apply a saddle point approximation to the resulting integral expression. This approximation does not directly rely on the geodesic approximation but shares the same regime of validity of large scalar masses.
\end{enumerate}
For equal-space correlators, we will find the same simple analytic expression in all three approaches.


\subsection{Continuing to Euclidean signature}\label{sect:Euclidean}

In appendix~\ref{app:geodesics}, we review how AdS and BTZ can be Wick rotated to Euclidean signature, and how various two-point correlation functions of high-dimension operators in the dual field theory can be computed using bulk geodesics.  Here we develop a similar approach for AdS-Vaidya.


\subsubsection{Wick rotation of AdS-Vaidya}

Recall that the thin-shell AdS-Vaidya metric
\begin{equation} \label{eq:Vaidyametric}
ds^{2}=- \left[r^{2}-\theta(v)R^{2} \right] dv^{2}+2dvdr+r^{2}dx^{2}
\end{equation}
represents an infalling shock wave of `null' dust in AdS$_{3}$. Outside the shock wave ($v>0$), the metric is given by the BTZ solution
\begin{equation} \label{eq:BTZmetric}
ds^{2}= -(r^{2}-R^{2})dt^{2} +\frac{dr^{2}}{r^{2}-R^{2}}+r^{2}dx^{2}
\qquad\text{with}\qquad
t=v-\frac{1}{2R}\ln\left(\frac{|r-R|}{r+R}\right)\,,
\end{equation}
while inside ($v<0$), we recover the pure AdS$_{3}$ metric
\begin{equation} \label{eq:AdSmetric}
ds^{2}=-r^{2}dt^{2} + \frac{dr^{2}}{r^{2}}+r^{2}dx^{2}
\qquad\text{with}\qquad
t=v+\frac{1}{r}.
\end{equation}
While for static spacetimes, such as  AdS$_{3}$ or BTZ, a Wick rotation of the metric is straightforward, this is no longer the case for an explicitly time dependent geometry.   Thus we propose the following procedure:
\begin{itemize}
\item  We  formally take the Vaidya metric to be a  limit of a family of ``spacelike Vaidya" metrics.    The latter need not be solutions to the equations of motion with conventional matter -- they are auxiliary objects used to regularize the computation.
\item This spacelike Vaidya metric can be Wick rotated to Euclidean signature by a simultaneous rotation of the time coordinate, as well as of an additional parameter.
\item In this  auxiliary Euclidean  spacetime, correlation functions can be computed in a geodesic approximation. The result can then be Wick rotated back.  Taking the limit we find our proposed result for the conventional AdS-Vaidya space-time.
\end{itemize}
While the results we find will look reasonable, it is not {\it a priori} obvious that this procedure  should be valid.  This is why, after discussing the present method in detail, we will also develop two alternative prescriptions.    All three prescriptions match, suggesting that the method of considering a time-dependent space-time as a limit of others with standard Euclidean continuations may be more generally useful.


\paragraph{Lorentzian `spacelike' Vaidya}\mbox{}\\

Let $E>R>0$ and consider the `spaceike' Vaidya metric
\begin{equation} \label{eq:spacelikeVaidya}
ds^{2}=-\left[r^{2}-\theta(v/E)R^{2}\right] dv^{2}+\frac{2\,E\,dv \, dr}{\sqrt{r^{2}+E^{2}-\theta(v/E)R^{2}}}+\frac{dr^{2}}{r^{2}+E^{2}-\theta(v/E)R^{2}}+r^{2}dx^{2}\,.
\end{equation}
Through the coordinate transformation
\begin{align}
v = \left\{\begin{array}{ll}
t-\frac{1}{R}\text{arctanh}\left(\frac{R}{E}\sqrt{1+\frac{(E^{2}-R^{2})}{r^{2}}}\right)+\frac{1}{R}\text{arctanh}\left(\frac{R}{E}\right) & {\rm for} \ v>0 \\
t+\frac{1}{E}\left(1-\sqrt{1+\frac{E^{2}}{r^{2}}}\right) & {\rm for} \ v<0
\end{array}\right.
\end{align}
we recover the BTZ metric \eqref{eq:BTZmetric} for $v>0$ and AdS$_3$ \eqref{eq:AdSmetric} for $v<0$.
The coordinate transformations are such that $v=0$ describes the geodesic of an infalling shell of `spacelike' particles in the BTZ and AdS$_{3}$ spacetimes. Observe that for $E>0$, $\theta(v/E)=\theta(v)$, so that in the $E\rightarrow\infty$ limit \eqref{eq:spacelikeVaidya} reduces to the Lorentzian `null' Vaidya metric \eqref{eq:Vaidyametric}.


\paragraph{Euclidean Vaidya}\mbox{}\\

On the Lorentzian `spacelike' Vaidya metric \eqref{eq:spacelikeVaidya}, we can now perform an analytic continuation, on the time coordinate $z=iv$, as well as on the parameter $Z=iE$. Without loss of generality, we can take $Z>0$ and find the metric
\begin{equation}
ds^{2}= \left[r^{2}-\theta(z/Z)R^{2} \right] dz^{2}-\frac{2 \,Z \,dz\, dr}{\sqrt{r^{2}-Z^{2}-\theta(z/Z)R^{2}}}+\frac{dr^{2}}{r^{2}-Z^{2}-\theta(z/Z)R^{2}}+r^{2}dx^{2}\,,
\end{equation}
where as before $\theta(z/Z)=\theta(z)$.
Note that the radial coordinate $r$ now runs from $\sqrt{Z^{2}+R^{2}}$ to $\infty$.

Letting
\begin{equation} \label{eq:finout}
f(r)^{2} \equiv \begin{cases}
f_{in}(r)^{2}=r^{2}-Z^{2} & z<0 \\
f_{out}(r)^{2}=r^{2}-Z^{2}-R^{2} & z>0
\end{cases}\,,
\end{equation}
the metric becomes
\begin{equation}
ds^{2}=f(r)^{2}dz^{2}+\left(Zdz-\frac{dr}{f(r)}\right)^{2}+r^{2}dx^{2}\,,
\end{equation}
which is manifestly positive.


\subsubsection{Geodesic length in Euclidean Vaidya}

We now compute the length of geodesics that start at a point $(r_{0},x_{1},y_{1})$ and end at $(r_{0},x_{2},y_{2})$, where $r_0$ denotes the location of the regularized AdS boundary. We assume $\Delta y \equiv y_{2}-y_{1}>0$ and, for computational simplicity, mostly focus on geodesics with $\Delta x \equiv x_{2}-x_{1}=0$.  Geodesics in the Euclidean Vaidya geometry can be constructed by gluing together at the shell, with the appropriate refraction condition, geodesics in Euclidean AdS and Euclidean BTZ. The latter have been worked out explicitly in Appendix~\ref{app:geodesics} and read
\begin{eqnarray}
x_{\pm}(r)&=&x_{0}\pm\frac{1}{R}\text{arctanh}\left(\frac{\Gamma_{-}}{\Gamma_{+}}\sqrt{\frac{r^{2}-\Gamma_{+}^{2}}{r^{2}-\Gamma_{-}^{2}}}\right), \nonumber \\
z_{\pm}(r)&=&y_{0}\pm\frac{1}{R}\text{arctan}\left(\sqrt{\frac{R^{2}-\Gamma_{-}^{2}}{\Gamma_{+}^{2}-R^{2}}}\sqrt{\frac{r^{2}-\Gamma_{+}^{2}}{r^{2}-\Gamma_{-}^{2}}}\right) \nonumber \\
&&\qquad\qquad\qquad+\frac{1}{R}\text{arctan}\left(\frac{R}{Z}\sqrt{1-\frac{(Z^{2}+R^{2})}{r^{2}}}\right)-\frac{1}{R}\text{arctan}\left(\frac{R}{Z}\right), \nonumber \\
\lambda_{\pm}(r)&=&\lambda_{0}\pm\text{arccosh}\left(\sqrt{\frac{r^{2}-\Gamma_{-}^{2}}{\Gamma_{+}^{2}-\Gamma_{-}^{2}}}\right)
\label{GeodesicsAboveShell}
\end{eqnarray}
for $z>0$, and
\begin{eqnarray}
\bar{x}_{\pm}(r)&=&\bar{x}_{0}\pm\frac{1}{\Sigma}\cos(\sigma)\sqrt{1-\frac{\Sigma^{2}}{r^{2}}}, \nonumber \\
\bar{z}_{\pm}(r)&=&\bar{y}_{0}\pm\frac{1}{\Sigma}\sin(\sigma)\sqrt{1-\frac{\Sigma^{2}}{r^{2}}}-\frac{1}{Z}+\frac{1}{Z}\sqrt{1-\frac{Z^{2}}{r^{2}}}, \nonumber \\
\bar{\lambda}_{\pm}(r)&=&\bar{\lambda}_{0}\pm\text{arccosh}\left(\frac{r}{\Sigma}\right)
\label{GeodesicsBelowShell}
\end{eqnarray}
for $z<0$. Here $\pm$ denote two separate branches (which appear above and below the turning points of AdS or BTZ geodesics) and $x_0$, $\Gamma_{\pm}$, $y_0$, $\lambda_0$, $\bar x_0$, $\Sigma$, $\sigma$, $\bar y_0$, $\bar \lambda_0$ are constants.

We can distinguish three cases.
\begin{itemize}
\item If both endpoints occur before shell injection  ($y_{1}<y_{2}<0$), there is a geodesic connecting them which is entirely within Euclidean AdS.   Any additional geodesics would require the existence of geodesics in BTZ that connect two points on the shell, and these do not exist.
\item   If both endpoints occur after shell injection  ($0<y_{1}<y_{2}$)), there is a geodesic connecting them which is entirely within Euclidean BTZ.\footnote{We can check that no new geodesics are admitted in the presence of the shell as follows. Consider  a geodesic with both endpoints in Euclidean BTZ, i.e. after shell injection ($0<y_{1}<y_{2}$). If the geodesic has to extend into the $z<0$ region, it will have to cross the shell twice.
It will therefore consist of two $z_{+}(r)$ branches, which are connected by a geodesic in AdS$_{3}$. $\bar{z}_{-}(r)$ thus needs to have a minimum, located at 
\begin{equation}
r_{\odot}=\frac{Z\Sigma\cos(\sigma)}{\sqrt{Z^{2}-\Sigma^{2}\sin^{2}(\sigma)}}.
\end{equation}
For the special case of geodesics that connect points on the boundary with $\Delta x=0$, one has $\cos(\sigma)=0$, such that $r_{\odot}=0$. However, since $r>\sqrt{Z^{2}+R^{2}}$, $\bar{z}_{-}(r)$ reaches no minimum and the geodesic is thus completely in Euclidean BTZ.  The renormalized geodesic length is worked out in Eq.~\eqref{eq:BTZlength} and reads
\begin{equation}\label{eq:lengthEBTZ}
\delta\mathcal{L}_{\rm EBTZ}=\ln\left[\frac{4}{R^{2}}\left(\sinh^{2}\left(\frac{R\Delta x}{2}\right)+\sin^{2}\left(\frac{R\Delta y}{2}\right)\right)\right]\,.
\end{equation}
}
\item Finally consider geodesics with $y_1<0$ and $y_2 >0$, and such that $\Delta x=0$. They cross the shock wave once and hence we need to determine how they refract at the shell location in order to compute their length.
\end{itemize}


\paragraph{Boundary conditions at the shell}\mbox{}\\

Consider a point $P_{in}$ ($P_{out}$) just inside (outside) the shell, and let the coordinate difference between $P_{in}$ and $P_{out}$ be $\Delta X=(\Delta z,\Delta r,\Delta t)$. Take another point $M$ on the shell $z=0$, and let the coordinate difference between $P_{in}$ ($P_{out}$) and $M$ be $\Delta X_{in}=(\Delta z_{in},\Delta r_{in},\Delta t_{in})$ ($\Delta X_{out}=(\Delta z_{out},\Delta r_{out},\Delta t_{out})$), so that $\Delta X_{in}+\Delta X_{out}=\Delta X$. Then the distance from $P_{in}$ to $P_{out}$ via $M$ is
\begin{eqnarray}
\Delta s&=&\sqrt{\Delta s_{in}^{2}}+\sqrt{\Delta s_{out}^{2}} \nonumber \\
&=&\sqrt{f_{in}^{2}\Delta z_{in}^{2}+\left(Z\Delta z_{in}-\frac{\Delta r_{in}}{f_{in}}\right)^{2}+r^{2}\Delta x_{in}^{2}} \nonumber \\
&&+\sqrt{f_{out}^{2}\Delta z_{out}^{2}+\left(Z\Delta z_{out}-\frac{(\Delta r-\Delta r_{in})}{f_{out}}\right)^{2}+r^{2}(\Delta x-\Delta x_{in})^{2}}\,,
\end{eqnarray}
where $f_{in}$ and $f_{out}$ where defined in  \eqref{eq:finout}.
We want to find the point $M$ that minimizes $\Delta s$. Extremizing this with respect to $\Delta r_{in}$ and $\Delta x_{in}$ we find
\begin{eqnarray}
\frac{1}{\sqrt{\Delta s_{in}^{2}}}\left(Z\Delta z_{in}-\frac{\Delta r_{in}}{f_{in}}\right)\frac{1}{f_{in}}&=&\frac{1}{\sqrt{\Delta s_{out}^{2}}}\left(Z\Delta z_{out}-\frac{\Delta r_{out}}{f_{out}}\right)\frac{1}{f_{out}}\nonumber \\
\frac{\Delta x_{in}}{\sqrt{\Delta s_{in}^{2}}}&=&\frac{\Delta x_{out}}{\sqrt{\Delta s_{out}^{2}}}
\end{eqnarray}
and therefore
\bea
\frac{1}{f_{in}}\left(Z\frac{\Delta z_{in}}{\Delta x_{in}}-\frac{1}{f_{in}}\frac{\Delta r_{in}}{\Delta x_{in}}\right)&=&\frac{1}{f_{out}}\left(Z\frac{\Delta z_{out}}{\Delta x_{out}}-\frac{1}{f_{out}}\frac{\Delta r_{out}}{\Delta x_{out}}\right) \nonumber\\
f_{in}^{2}\left(\frac{\Delta z_{in}}{\Delta x_{in}}\right)^{2}+\left(Z\frac{\Delta z_{in}}{\Delta x_{in}}-\frac{1}{f_{in}}\frac{\Delta r_{in}}{\Delta x_{in}}\right)^{2}&=&f_{out}^{2}\left(\frac{\Delta z_{out}}{\Delta x_{out}}\right)^{2}+\left(Z\frac{\Delta z_{out}}{\Delta x_{out}}-\frac{1}{f_{out}}\frac{\Delta r_{out}}{\Delta x_{out}}\right)^{2} \,. \nonumber\\
&&
\eea
In the $\Delta X\rightarrow0$ limit, we obtain the refraction law
\bea\label{eq:refractionlaw}
\left. \frac{1}{f^2 \frac{dx}{dr}} \left( Z f \frac{d z}{dr} -1 \right)\right|_{in} &=& \left. \frac{1}{f^2 \frac{dx}{dr}} \left( Z f \frac{d z}{dr} -1 \right)\right|_{out}\\
\left. \frac{1}{ \left(f \frac{dx}{dr}\right)^2} \left[ f^4 \left(\frac{dz}{dr} \right)^2 + \left( Z f \frac{dz}{dr} -1\right)^2 \right]\right|_{in} &=& \left. \frac{1}{ \left(f \frac{dx}{dr}\right)^2} \left[ f^4 \left(\frac{dz}{dr} \right)^2 + \left( Z f \frac{dz}{dr} -1\right)^2 \right]\right|_{out} \nonumber
\eea
that is supplemented by the continuity conditions $x_{in}(r_{*})=x_{out}(r_{*})$ and $z_{in}(r_{*})=z_{out}(r_{*})=0$, where $r_*$ is the value of the radial coordinate at which the geodesic reaches the shell.


\paragraph{Geodesics crossing the shell}\mbox{}\\

If $y_{1}<0<y_{2}$, the geodesic starts in Euclidean AdS, crosses the shell once and ends in Euclidean BTZ. The geodesics are given respectively by the expressions (\ref{GeodesicsBelowShell}) and (\ref{GeodesicsAboveShell}). Imposing the boundary conditions results in the following relations
\begin{eqnarray}
\begin{cases}
x_{2}=x_{0}+\frac{1}{R}\text{arctanh}\left(\frac{\Gamma_{-}}{\Gamma_{+}}\right) \\
y_{2}=y_{0}+\frac{1}{R}\text{arctan}\left(\sqrt{\frac{(R^{2}-\Gamma_{-}^{2})}{(\Gamma_{+}^{2}-R^{2})}}\right) \\
x_{1}=\bar{x}_{0}-\frac{1}{\Sigma}\cos(\sigma) \\
y_{1}=\bar{y}_{0}-\frac{1}{\Sigma}\sin(\sigma)
\end{cases}
\end{eqnarray}
Together with the refraction condition \eqref{eq:refractionlaw} that leads to
\bea
R^{2}\Sigma^{2}\cos^{2}(\sigma) &=& \Gamma_{+}^{2}\Gamma_{-}^{2}\\
\left(\frac{Z\Sigma\sin(\sigma)}{\sqrt{r_{*}^{2}-Z^{2}}}-\sqrt{r_{*}^{2}-\Sigma^{2}}\right) &=&  \frac{r_{*}}{(r_{*}^{2}-R^{2})}\left(\frac{Zr_{*}\sqrt{(\Gamma_{+}^{2}-R^{2})(R^{2}-\Gamma_{-}^{2})}}{R\sqrt{r_{*}^{2}-(Z^{2}+R^{2})}}-\sqrt{(r_{*}^{2}-\Gamma_{+}^{2})(r_{*}^{2}-\Gamma_{-}^{2})}\right) \nonumber
\eea
and the continuity conditions at the shell
\begin{align}
&\bar{x}_{0}+\frac{1}{\Sigma}\cos(\sigma)\sqrt{1-\frac{\Sigma^{2}}{r_{*}^{2}}}=x_{0}+\frac{1}{R}\text{arctanh}\left(\frac{\Gamma_{-}}{\Gamma_{+}}\sqrt{\frac{r_{*}^{2}-\Gamma_{+}^{2}}{r_{*}^{2}-\Gamma_{-}^{2}}}\right),\\
&\bar{y}_{0}+\frac{1}{\Sigma}\sin(\sigma)\sqrt{1-\frac{\Sigma^{2}}{r_{*}^{2}}}-\frac{1}{Z}+\frac{1}{Z}\sqrt{1-\frac{Z^{2}}{r_{*}^{2}}} =0 \\
&y_{0}+\frac{1}{R} \left[\text{arctan}\left(\sqrt{\frac{R^{2}-\Gamma_{-}^{2}}{\Gamma_{+}^{2}-R^{2}}}\sqrt{\frac{r_{*}^{2}-\Gamma_{+}^{2}}{r_{*}^{2}-\Gamma_{-}^{2}}}\right)+ \text{arctan}\left(\frac{R}{Z}\sqrt{1-\frac{(Z^{2}+R^{2})}{r_{*}^{2}}}\right)-\text{arctan}\left(\frac{R}{Z}\right) \right]= 0
\end{align}
these fix all unknowns ($x_{0}$, $y_{0}$, $\bar{x}_{0}$, $\bar{y}_{0}$, $\Sigma$, $\sigma$, $\Gamma_{+}$, $\Gamma_{-}$ and $r_{*}$) and determine the renormalized geodesic length
\bea \label{eq:lengthEShell1}
\delta\mathcal{L}_{\rm EShell} &\equiv& \lim_{r_{0}\rightarrow\infty}\left[\lambda_{+}(r_{0})-\bar{\lambda}_{-}(r_{0})-2\ln(r_{0})\right]\nonumber \\
&=&\lambda_{0}-\bar{\lambda}_{0}+\frac{1}{2}\ln\left(\frac{4}{\Sigma^{2}}\right)+\frac{1}{2}\ln\left(\frac{4}{\Gamma_{+}^{2}-\Gamma_{-}^{2}}\right) \nonumber \\
&=&-\text{arctanh}\left(\sqrt{\frac{r_{*}^{2}-\Gamma_{+}^{2}}{r_{*}^{2}-\Gamma_{-}^{2}}}\right)+\text{arctanh}\left(\sqrt{1-\frac{\Sigma^{2}}{r_{*}^{2}}}\right)+\frac{1}{2}\ln\left(\frac{4}{\Sigma^{2}}\right)+\frac{1}{2}\ln\left(\frac{4}{\Gamma_{+}^{2}-\Gamma_{-}^{2}}\right)\,,\nonumber \\
&&
\eea
where in the last line we have used the continuity condition $\lambda_{+}(r_{*})=\bar{\lambda}_{+}(r_{*})$.

When $\Delta x=0$: $x_{0}=\bar{x}_{0}=x_{1}=x_{2}$, $\sigma=\pi/2$ and $\Gamma_{-}=0$,  the geodesic length \eqref{eq:lengthEShell1} becomes
\begin{equation}\label{eq:lengthEShell2}
\delta\mathcal{L}_{\rm EShell}=\frac{1}{2}\ln\left\{ \frac{r_{*}-\cos(\gamma(r_{*}))\sqrt{r_{*}^{2}-R^{2}}}{r_{*}+\cos(\gamma(r_{*}))\sqrt{r_{*}^{2}-R^{2}}} \frac{4\left[(r_{*}M(r_{*}))^{2}+1\right]^{2}}{r_{*}^{2}\left[(r_{*}^{2}-R^{2})\sin^{2}(\gamma(r_{*}))+R^{2}\right]}\right\},
\end{equation}
where $r_{*}$ is defined by the parametric relation
\begin{equation}
\frac{1}{(r_{*}M(r_{*}))^{2}+1}\left[\frac{2Zr_{*}M(r_{*})}{\sqrt{r_{*}^{2}-Z^{2}}}-(r_{*}M(r_{*}))^{2}+1\right]=\frac{r_{*}}{\sqrt{r_{*}^{2}-R^{2}}}\left[\frac{Z\sin(\gamma(r_{*}))}{\sqrt{r_{*}^{2}-(Z^{2}+R^{2})}}-\cos(\gamma(r_{*}))\right],
\end{equation}
and
\bea
M(r_{*}) &\equiv& -y_{1}+\frac{1}{Z}-\frac{1}{Z}\sqrt{1-\frac{Z^{2}}{r_{*}^{2}}}\,, \\
\sin(\gamma(r_{*})) & \equiv& \frac{R\sin(R\Omega(r_{*}))}{r_{*}-\cos(R\Omega(r_{*}))\sqrt{r_{*}^{2}-R^{2}}}\,,
\eea
with
\begin{equation}
\Omega(r_{*}) \equiv y_{2}+\frac{1}{R}\text{arctan}\left(\frac{R}{Z}\sqrt{1-\frac{(Z^{2}+R^{2})}{r_{*}^{2}}}\right)-\frac{1}{R}\text{arctan}\left(\frac{R}{Z}\right)\,.
\end{equation}

Performing the double Wick rotation $y_{1}=it_{1}$, $y_{2}=it_{2}$ and $Z=iE$ of \eqref{eq:lengthEShell2} and taking the $E\to \infty$ limit, in order to recover the `lightlike' Vaidya result, we finally find
\begin{equation}
\delta\mathcal{L}_{\rm Shell}=  \ln\left[ \left( \frac{2}{R}\sinh\left(\frac{Rt_{2}}{2}\right)-\cosh\left(\frac{Rt_{2}}{2}\right)t_{1} \right)^2\right].
\label{GeodesicLengthVaidya}
\end{equation}


\subsubsection{Two-point functions of thermalising CFT}\label{sect:spectralgeodesic}

 As reviewed in appendix~\ref{app:geodesics} for the cases of AdS and BTZ, by Wick rotating the Euclidean two point function $G_E(x_2,y_2; x_1, y_1)$, one finds the time-ordered (Feynman) two-point function\footnote{The $\epsilon$ prescriptions are given in Appendix~\ref{integerminkowski}.}
\begin{equation}
iG_F (x_{2},t_{2};x_{1},t_{1})=G_E(x_{2},it_{2};x_{1},it_{1})\,.
\end{equation}
The retarded two-point function can be obtained from the time-ordered one by
\begin{equation}
G_R(x_{2},t_{2};x_{1},t_{1})=\theta(t_{2}-t_{1})\left[G_F(x_{2},t_{2};x_{1},t_{1})+(G_F(x_{2},t_{2};x_{1},t_{1}))^{*}\right].
\end{equation}

From \eqref{GeodesicLengthVaidya} we then find (the geodesic approximation of) the Feynman and retarded two-point functions in the thermalizing CFT dual to 3-dimensional AdS-Vaidya:
\begin{eqnarray}
i G_F (t_{2},x;t_{1},x)&=&\left(\frac{e^{-i\pi\Delta}}{|t_{2}-t_{1}|^{2\Delta}}\right)\theta(-t_{2})\theta(-t_{1})+\left(\frac{e^{-i\pi\Delta}}{\left|\frac{2}{R}\sinh\left(\frac{R}{2}(t_{2}-t_{1})\right)\right|^{2\Delta}}\right)\theta(t_{2})\theta(t_{1}) \nonumber \\
&&+\left(\frac{e^{-i\pi\Delta}}{\left|\frac{2}{R}\sinh\left(\frac{Rt_{2}}{2}\right)-\cosh\left(\frac{Rt_{2}}{2}\right)t_{1}\right|^{2\Delta}}\right)\theta(t_{2})\theta(-t_{1}) \nonumber \\
&&+\left(\frac{e^{-i\pi\Delta}}{\left|\frac{2}{R}\sinh\left(\frac{Rt_{1}}{2}\right)-\cosh\left(\frac{Rt_{1}}{2}\right)t_{2}\right|^{2\Delta}}\right)\theta(-t_{2})\theta(t_{1})\,, \\
G_R (t_{2},x;t_{1},x)&=&-2\sin(\pi\Delta)\theta(t_{2}-t_{1})\left(\frac{\theta(-t_{2})\theta(-t_{1})}{|t_{2}-t_{1}|^{2\Delta}}+\frac{\theta(t_{2})\theta(t_{1})}{\left|\frac{2}{R}\sinh\left(\frac{R}{2}(t_{2}-t_{1})\right)\right|^{2\Delta}}\right. \nonumber \\
&&\qquad\qquad\qquad\qquad\qquad\left.+\frac{\theta(t_{2})\theta(-t_{1})}{\left|\frac{2}{R}\sinh\left(\frac{Rt_{2}}{2}\right)-\cosh\left(\frac{Rt_{2}}{2}\right)t_{1}\right|^{2\Delta}}\right). \label{eq:GRVaidya}
\end{eqnarray}
In the limit $R\rightarrow0$ or  $t_2 \to 0^+$, $G_R (t_{2},x;t_{1},x)$ reduces to the retarded vacuum two-point function \eqref{eq:GRvacuum} with $\Delta t = t_2 -t_1$ or $\Delta t = -t_1$ respectively, while for $t_1 \to 0^-$ to the thermal correlator \eqref{eq:GRthermal} with $\Delta t = t_2$.   The above formulas are valid for non-integer values of $\Delta$.  For a discussion of integer values of $\Delta$, see Appendix C.

To compute time-dependent spectral functions and occupation numbers, we would need to Fourier transform the relative times and positions in the retarded and time-ordered two-point functions. This requires knowledge of the position space two-point functions for arbitrary separations, which are more cumbersome to obtain than the equal-space two-point functions we have explicitly computed so far. We will therefore focus on the information we can extract from the equal-space two-point functions, {\it i.e.} the time-dependent spectral function integrated over all spatial momenta.

Following the discussion in Section~\ref{spectral}, we introduce relative and average time coordinates $t \equiv t_{2}-t_{1}$ and $T \equiv (t_1+t_2)/2$ and Fourier transform $G_R(t_{2},x;t_{1},x)$ with respect to $t$. The imaginary part of the result is
\begin{equation}
\rho(T,\omega)=\int_{-\infty}^{+\infty} \text{d}k \,  \rho(k,T, \omega)=  -4\pi\,\text{Im}\left(\int_{-\infty}^{+\infty}\text{d}t\,e^{i\omega t}G_R(x=0,T,t)\right)\,,
\end{equation}
where the additional factor of $2\pi$ with respect to the definition \eqref{eq:rhoT}  appears because the retarded two point function is here given in position space, rather than momentum space.

\begin{figure}[h]
\begin{center}
\includegraphics[width=0.45 \textwidth]{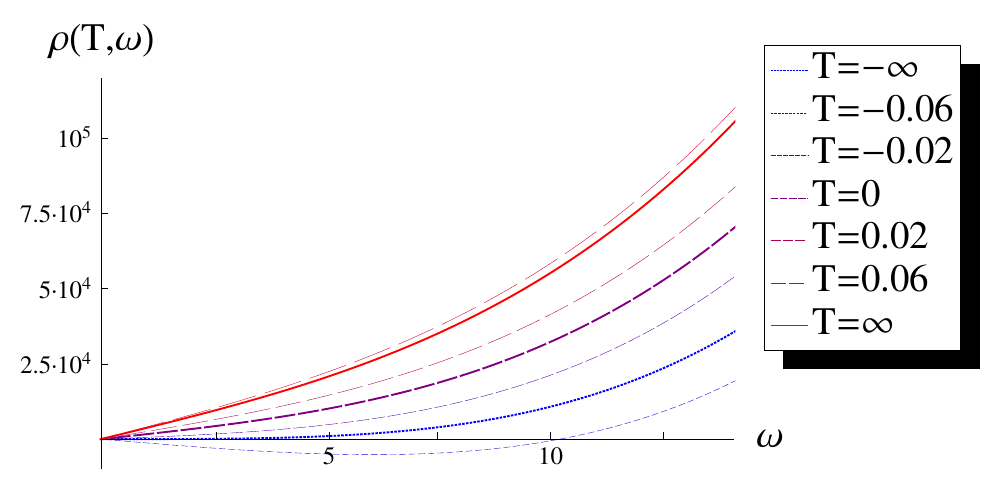}
\hfil
\includegraphics[width=0.45 \textwidth]{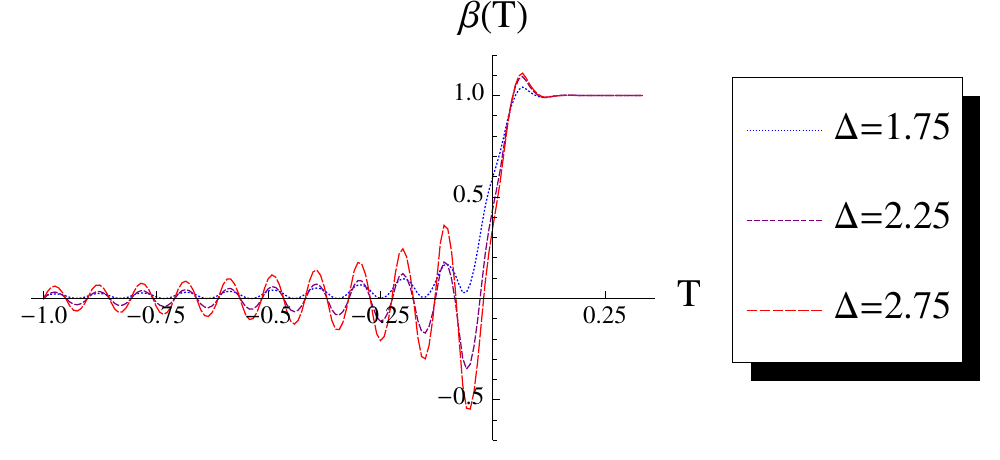}\\
{\bf (A)} \hfil {\bf (B)}
\caption{Panel {\bf (A)} plots the numerical results for the spectral function integrated over momenta $\rho(T,\omega)$ for different values of the mean time $T$ (with $\Delta=9/4$, $R=15$). When $T\ll -1/R$, the curve reduces to the the vacuum result (in blue dotted), while for $T\gg 1/R$, to the thermalised one (in red continuous). (The reference case $T=0$ is also shown in purple.) Close to the origin, the spectral function grows linearly in  frequency. The slopes $\beta$ of this regime of linear growth are plotted in panel {\bf (B)} as a function of time and for various values of $\Delta$.  Similar results are obtained for larger values of $\Delta$. The oscillations in the slope characterize how the spectral function interpolates between the vacuum and thermal results.   
}
\label{fig:spectralfunction}
\end{center}
\end{figure}

Numerical results for the spectral function integrated over momenta are plotted in Fig.~\ref{fig:spectralfunction} for different times. As time evolves, the curves interpolate between the vacuum result (in dotted blue in panel {\bf (A)}) and the thermal one (in continuous red). As the spectral function is odd in $\omega$, we only plot the range $\omega \ge 0$. Panel {\bf (B)} plots, as a function of time and for different values of scaling dimension $\Delta$, the slope $\beta$ of the spectral function in the range of linear growth close to $\omega =0$. For a fixed frequency $\omega$,  $\rho(\omega, T)$ does not increase monotonically from the vacuum to the thermal value, but it oscillates.

\begin{figure}[h]
\begin{center}
\includegraphics[width=0.45 \textwidth]{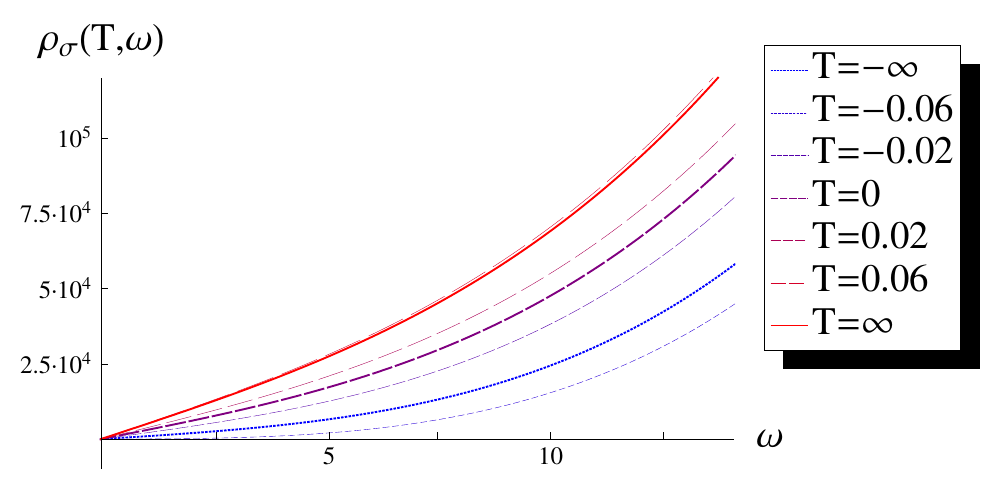}
\hfil
\includegraphics[width=0.45 \textwidth]{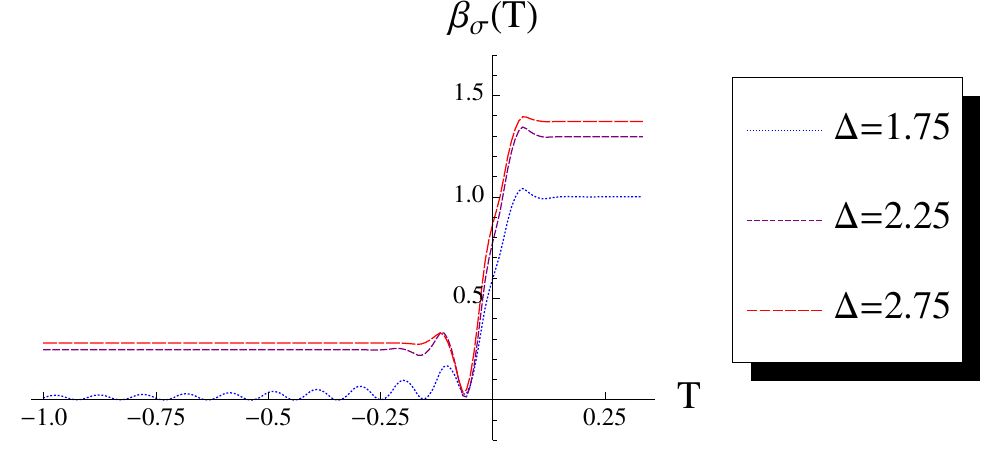}\\
{\bf (A)} \hfil {\bf (B)}
\caption{Spectral function integrated over momenta {\bf (A)} and slopes of the regime of linear growth  {\bf (B)} as in Fig.~\ref{fig:spectralfunction}, with time window set by $\sigma=2.9/R$.
}
 \label{fig:spectralfunctionsigma}
\end{center}
\end{figure}

As in the example of the quenched harmonic oscillator we discussed in Sect.~\ref{harmonic}, the time-dependent spectral function we have defined can be negative for positive $\omega$. Introducing an appropriate time window as in Eq.~\eqref{eq:Gsigma}-\eqref{eq:rhosigma}, we can smear the curves and make the spectral function positive definite for $\omega >0$ (see Fig.~\ref{fig:spectralfunctionsigma}).

\begin{figure}[h]
\begin{center}
\includegraphics[width=0.45 \textwidth]{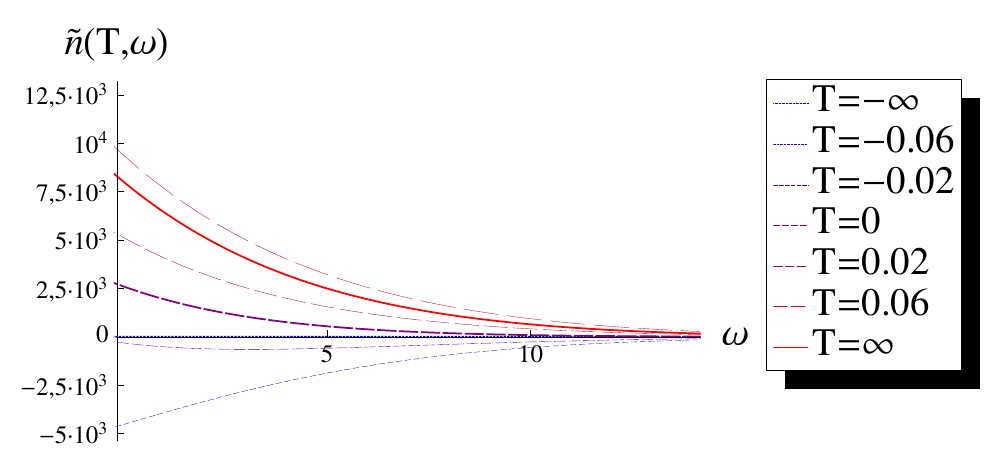}
\hfil
\includegraphics[width=0.45 \textwidth]{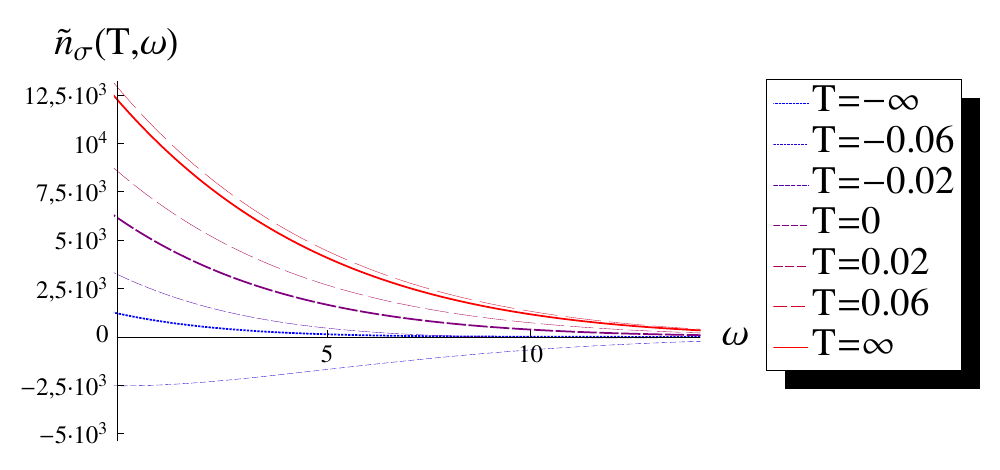}\\
{\bf (A)} \hfil {\bf (B)}
\caption{{\bf (A)} The integrated occupation number weighted with the spectral function. The times for which this quantity becomes negative corresponds to the same times for which the spectral function has a region that satisfies $\omega\rho(\omega,T)<0$. {\bf (B)} The same quantity with a time-window set by $\sigma=2.9/R$.
}
 \label{fig:occupationnumbers}
\end{center}
\end{figure}

From the time-ordered two-point function, using definition \eqref{eq:nsigma}, we can also obtain time-dependent occupation numbers weighted by the spectral function and integrated over all spatial momenta $\tilde n(T,\omega)$. These are plotted in Fig.~\ref{fig:occupationnumbers} as a function of time, without (in {\bf (A)}) and with (in {\bf (B)}) the Gaussian smearing.
%


\subsection{Complexified geodesics}\label{sect:complex}

To investigate the validity of the non standard Euclidean continuation used in Sect.~\ref{sect:Euclidean}, we here repeat the computation of the geodesic length in the 3-dimensional AdS-Vaidya spacetime using a different approach. While real timelike geodesics can not connect timelike separated points on the boundary, complex geodesics can. The latter have been studied in \cite{Kraus:2002iv,Fidkowski:2003nf,Festuccia:2005pi} and we review their construction in both AdS$_3$ and BTZ in Appendix~\ref{app:complex}. As we show below, the geodesic length obtained with this second method coincides with (\ref{GeodesicLengthVaidya}).


As before, we can construct Vaidya geodesics that cross the shockwave by joining together AdS and BTZ geodesics, with an appropriate refraction condition at the shell. For equal space geodesics, the relevant formulae are given in Eq.~\eqref{eq:complexAdS} and \eqref{eq:complexBTZ}  with $j=J=0$. For $v>0$, the coordinate transformation $v=t-\frac{1}{R}\text{arccoth}\left(\frac{r}{R}\right)$ leads to
\begin{eqnarray}
r_{2}(\lambda)&=&\sqrt{E^{2}-R^{2}}\sinh(\lambda-\lambda_{2}+i\beta_{2}) \\
t_{2}(\lambda)&=&\tau_{2}-\frac{1}{R}\text{arccoth}\left[\frac{E}{R}\tanh(\lambda-\lambda_{2}+i\beta_{2})\right] \\
v_{2}(\lambda)
&=&\tau_{2}-\frac{2}{R}\text{arccoth}\left[\frac{E+\sqrt{E^{2}-R^{2}}}{R}\tanh\left(\frac{1}{2}(\lambda-\lambda_{2}+i\beta_{2})\right)\right],
\end{eqnarray}
while for $v<0$, $v$ is defined as $v=t-\frac{1}{r}$ and the geodesics are given by
\begin{eqnarray}
r_{1}(\lambda)&=&e\sinh(\lambda-\lambda_{1}+i\beta_{1}) \\
t_{1}(\lambda)&=&\tau_{1}-\frac{1}{e}\coth(\lambda-\lambda_{1}+i\beta_{1}) \\
v_{1}(\lambda)&=&\tau_{1}-\frac{1}{e}\coth\left[\frac{1}{2}(\lambda-\lambda_{1}+i\beta_{1})\right].
\end{eqnarray}
The refraction condition at the shell has been computed in \cite{Balasubramanian:2010ce,Balasubramanian:2011ur} and reads
\bea
&&\left.\frac{r'(\lambda_{*})}{v'(\lambda_{*})}\right|_{\text{in}}-\left.\frac{r'(\lambda_{*})}{v'(\lambda_{*})}\right|_{\text{out}}=\frac{R^{2}}{2}\,,\\
&& \left.\frac{x'(\lambda_{*})}{v'(\lambda_{*})}\right|_{\text{in}}-\left.\frac{x'(\lambda_{*})}{v'(\lambda_{*})}\right|_{\text{out}}=0\,,
\eea
where the second condition is trivially satisfied for equal-space geodesics.
Continuity at the shell fixes
\begin{align}
e\sinh(\Lambda_{1}) =&\sqrt{E^{2}-R^{2}}\sinh(\Lambda_{2}) \label{eq:continuity1} \\
 \tau_{2} =& \frac{2}{R}\text{arccoth}\left[\frac{E+\sqrt{E^{2}-R^{2}}}{R}\tanh\left(\frac{\Lambda_{2}}{2}\right)\right]  \\
\tau_{1}=&\frac{1}{e}\coth\left(\frac{\Lambda_{1}}{2}\right)  \,,
\end{align}
where we have defined $\Lambda_{1} \equiv \lambda_{*}-\lambda_{1}+i\beta_{1}$ and $\Lambda_{2} \equiv \lambda_{*}-\lambda_{2}+i\beta_{2}$. The boundary conditions are
\begin{align}
t_{2}&=\tau_{2}-\frac{2}{R}\text{arccoth}\left(\frac{E+\sqrt{E^{2}-R^{2}}}{R}\right) \\
t_{1}&=\tau_{1}+\frac{1}{e}
\end{align}
 and using the continuity condition \eqref{eq:continuity1},
the refraction condition can be written as
\begin{equation}
\frac{R^{2}}{2}=E\sqrt{E^{2}-R^{2}}\cosh(\Lambda_{2})-e^{2}\cosh(\Lambda_{1})+e^{2}-E^{2}+R^{2}\,.
\end{equation}
Finally, we find the solution
\begin{equation}
E=\frac{R\left[\cosh\left(\frac{Rt_{2}}{2}\right)-\frac{Rt_{1}}{2}\sinh\left(\frac{Rt_{2}}{2}\right)\right]}{\frac{Rt_{1}}{2}\cosh\left(\frac{Rt_{2}}{2}\right)-\sinh\left(\frac{Rt_{2}}{2}\right)}
\qquad\text{and}\qquad
e=\frac{R\cosh\left(\frac{Rt_{2}}{2}\right)}{\frac{Rt_{1}}{2}\cosh\left(\frac{Rt_{2}}{2}\right)-\sinh\left(\frac{Rt_{2}}{2}\right)}
\end{equation}
\begin{equation}
\exp ({\Lambda_{1}})=\frac{Rt_{1}}{2}\coth\left(\frac{Rt_{2}}{2}\right)
\qquad\text{and}\qquad
\exp ({-\Lambda_{2}})=\sinh\left(\frac{Rt_{2}}{2}\right)\sqrt{\left(\frac{2}{Rt_{1}}\right)^{2}-1},
\end{equation}
such that
\begin{eqnarray}
\delta\mathcal{L}_{\rm Shell }&=&\frac{1}{2}\ln\left(\frac{4}{e^{2}}\right)+\frac{1}{2}\ln\left(\frac{4}{E^{2}-R^{2}}\right)+\lambda_{2}-\lambda_{1}=\frac{1}{2}\ln\left(\frac{16e^{2\Lambda_{2}}e^{-2\Lambda_{1}}}{e^{2}(E^{2}-R^{2})}\right) \nonumber \\
&=&\ln\left[\left(\frac{2}{R}\sinh\left(\frac{Rt_{2}}{2}\right)-\cosh\left(\frac{Rt_{2}}{2}\right)t_{1}\right)^{2}\right],
\end{eqnarray}
which corresponds exactly to the result \eqref{GeodesicLengthVaidya} that we found using the analytic continuation to Euclidean signature.


\subsection{A saddle point approximation for joining propagators}\label{sect:joining}

In this section we consider a third way of calculating the equal space retarded propagator in the AdS-Vaidya background.
Specifically we will consider the saddle point approximation, which is valid in the limit of large scaling dimension $\Delta$, of the following integral formula for the CFT correlation function
\begin{align}
G_R(x_2, t_2;&x_1,t_1)=i\int_{v=0} dx\,d z\sqrt{-g}g^{v\mu}\Big[G_{B,R}^{\rm AdS}(x,v=0,z;x_1,t_1)\overleftrightarrow{\partial_{\mu}}\tilde{G}_{B,R}^{\rm BTZ}(x_2,t_2;x,v=0,z)\Big],
\label{eq:mainformula}
\end{align}
where $t_1 <0 < t_2$, and $G_{B,R}$ and $\tilde{G}_{B,R}$ are propagators that we will describe in more detail below.   First, we have switched to coordinates $(x,v,z)$ in which the AdS-Vaidya metric reads
\be \label{eq:Vaidyaz}
ds^2 = \frac{1}{z^2} \left[ -\left( 1- \theta(v)R^2 z^2\right) dv^2 - 2 dv dz +dx^2\right]\,.
\ee
The time coordinate $t$ appearing in \eqref{eq:mainformula} is the boundary value of the Eddington-Finkelstein time $v$.

In Lorentzian signature one has to distinguish between bulk-to-boundary propagators and boundary-to-bulk propagators. The retarded boundary-to-bulk propagator $G_{B,R}$ having its initial point at the boundary and the final point in the bulk, evolves initial data from the boundary into the bulk.  On the other hand, the retarded bulk-to-boundary propagator $\tilde{G}_{B,R}$ takes bulk data and evolves it towards the spacetime boundary. Below we give a precise definition of these two-point functions in terms of the bulk-to-bulk propagator.

In (\ref{eq:mainformula}), the two sided derivative is understood to act only inside the square brackets,  $G_{B,R}^{\rm AdS}$ is the retarded boundary-to-bulk propagator in AdS$_3$ and $\tilde{G}_{B,R}^{\rm BTZ}$ is the retarded bulk-to-boundary propagator in BTZ. The integral in (\ref{eq:mainformula}) is performed at the position of the shell $v=0$ and $g$ is either the BTZ or the AdS$_3$ metric.\footnote{For the above formula to hold it is necessary for $\sqrt{-g}g^{v\mu}$ to be continuous at the junction $v=0$. As this is the case in the Vaidya spacetime, using either the BTZ or the AdS$_3$ metric gives the same result.}   The basic logic is as follows.  A delta-function source inserted on the boundary before the shell injection produces a disturbance that propagates to the shell according to $G_{B,R}^{AdS}$.   After passing through the shell, the disturbance propagates back to the boundary according to $\tilde{G}_{B,R}^{BTZ}$.   In the following we provide a derivation of (\ref{eq:mainformula}).


\paragraph{Derivation of Eq.~\ref{eq:mainformula}: }
First consider the retarded bulk-to-bulk propagator $G_{BB,R}$ associated to a scalar field of mass $m$ in an asymptotically AdS spacetime. It solves the equations
\begin{align}
(\Box-m^2)G_{BB,R}(x',v',z';x,v,z)&=(\Box'-m^2)G_{BB,R}(x',v',z';x,v,z)\nonumber
\\
&=-i\frac{\delta(v-v')\delta(x-x')\delta(z-z')}{\sqrt{-g}}\,,\label{eq:bulktobulk}
\end{align}
vanishes when $v'<v$ and satisfies the standard normalizable boundary condition at the AdS boundary.\footnote{Here we focus on scalar fields satisfying the standard normalizable boundary conditions $\Phi\propto z^{\Delta}$ as $z\rightarrow 0$, but generalizations to more general boundary conditions are in principle straightforward.}

Given the initial data $\Phi(x',v'=0,z')=\Phi_0(x',z')$, for $v' > 0$ we will show that the unique solution to the massive wave equation satisfying the boundary condition of vanishing non-normalizable mode is given by
\beq
\Phi(x',v',z')=i\int_{v=0}  dx \, dz \sqrt{-g}g^{v\mu}\Big[\Phi_0(x,z)\overleftrightarrow{\partial_\mu} G_{BB,R}(x',v',z';x,v=0,z)\Big]\,.\label{eq:wavesol}
\eeq
For $v'>0$, it is clear from the definition of the bulk-to-bulk propagator \eqref{eq:bulktobulk} that (\ref{eq:wavesol}) solves the
massive wave equation. By integrating (\ref{eq:bulktobulk}) with respect to $v$ from $v'-\epsilon$ to $v'+\epsilon$ we get
\begin{align}
\int_{v'-\epsilon}^{v'+\epsilon}dv&\Big(\partial_{\mu}(\sqrt{-g}g^{\mu\nu}\partial_{\nu}G_{BB,R}(x',v',z';x,v=0,z))-m^2\sqrt{-g} G_{BB,R}(x',v',z';x,v=0,z)\Big)\nonumber
\\
&=-i\delta(x-x')\delta(z-z').\label{eq:matchingintegral}
\end{align}
The equations of motion (\ref{eq:bulktobulk}) imply that $G_{BB,R}$ is a regular function of $v$ near $v=v'$. Thus, the mass term as well as the $\partial_x(\sqrt{-g}g^{xx}\partial_x G_{BB,R})$ term can be assumed to be regular near $v=v'$ leading to order $\epsilon$ contributions to the integral in (\ref{eq:matchingintegral}). The terms singular near $v=v'$ in the integrand come from the derivative terms
\beq
\partial_v(\sqrt{-g}g^{vz}\partial_z G_{BB,R})+\partial_z(\sqrt{-g}g^{zv}\partial_v G_{BB,R}),
\eeq
which can be written as
\beq
\partial_v(\sqrt{-g}g^{vz}\partial_z G_{BB,R}+\partial_z(\sqrt{-g}g^{zv}G_{BB,R}))-\partial_z(\partial_v(\sqrt{-g}g^{zv})G_{BB,R})\,,
\eeq
where we have dropped the explicit coordinate dependence.
Using the fact that $\sqrt{-g}g^{vz}$ is a continuous function of $v$, and assuming that $\partial_z G_R$ is a regular function of $v$,\footnote{This assumption is indeed consistent with the final result we obtain.} the last term above gives a contribution of order $\epsilon$ to the integral in (\ref{eq:matchingintegral}). Finally the integral (\ref{eq:matchingintegral}) can be performed to give
\beq
\Big[\sqrt{-g}g^{vz}\partial_{z}G_{BB,R}+\partial_{z}(\sqrt{-g}g^{zv}G_{BB,R})\Big]_{v=v'-\epsilon}=-i\delta(x-x')\delta(z-z'),
\eeq
where we used the fact that $G_R$ vanishes when $v>v'$. Thus, we see that the bulk-to-bulk propagator satisfies
\begin{align}
\sqrt{-g}g^{v\mu}\partial_{\mu}&G_{BB,R}(x',v',z';x,v=0,z)+\partial_{\mu}(\sqrt{-g}g^{v\mu}G_{BB,R}(x',v',z';x,v=0,z))\nonumber
\\
&\rightarrow-i\theta(v')
\delta(x-x')\delta(z-z'),
\end{align}
as $v' \rightarrow 0$; so that indeed (\ref{eq:wavesol}) satisfies the correct initial condition $\Phi(x,v=0,z)=\Phi_0(x,z)$.

In particular we can use (\ref{eq:wavesol}) to join two retarded bulk-to-bulk propagators
\begin{align}
G_{BB,R}(x_2,v_2,z_2;x_1,v_1,z_1)&=i \int_{v=0} dx\, dz \sqrt{-g}g^{v\mu}\Big[G_{BB,R}( x,v=0, z;x_1v_1, ,z_1)\times\nonumber
\\
&\times \overleftrightarrow{\partial_\mu} G_{BB,R }(x_2,v_2,z_2;x,v=0,z)\Big] \,. \label{eq:bulktobulk2}
\end{align}
To obtain the boundary theory retarded correlation function $G_R$ from a bulk-to-bulk propagator we can use the well known relation (see e.g. \cite{Banks:1998dd,Festuccia:2005pi})
\beq
G_{R}( x_2,t_2;  x_1,t_1)=\lim_{z_1,z_2\rightarrow 0}(2\nu)^2 (z_1 z_2)^{-\Delta} G_{BB,R }(x_2,v_2,z_2; x_1,v_1,z_1)\,,
\eeq
where $t$ is the boundary value of $v$.
From (\ref{eq:bulktobulk2}) we then obtain
\beq
G_{R}(x_2,t_2;x_1,t_1)=i\int_{v =0} dx \,dz \sqrt{-g}g^{v\mu}G_{B,R}(x,v=0,z;x_1,t_1) \overleftrightarrow{\partial_\mu} \tilde{G}_{B,R}(x_2,t_2;x,v=0,z)\,,\label{eq:boundarytoboundary}
\eeq
where we have identified the retarded boundary-to-bulk propagator as \cite{Banks:1998dd}
\beq \label{eq:Gbdrybulk}
G_{B,R}(x,v=0,z; x_1,t_1)=2\nu\lim_{z_1 \rightarrow 0}z_1^{-\Delta}G_{BB,R}(x,v=0,z;x_1 ,v_1 ,z_1)\,,
\eeq
and the retarded bulk-to-boundary propagator as
\beq\label{eq:Gbulkbdry}
\tilde{G}_{B,R}(x_2,t_2;x,v=0,z)=2\nu\lim_{z_2\rightarrow 0}z_2^{-\Delta}G_{BB,R,}(x_2,v_2, z_2;x,v=0,z)\,.
\eeq


\paragraph{Application to AdS-Vaidya: } In the specific case of AdS$_3$ Vaidya,  \eqref{eq:boundarytoboundary} reduces to \eqref{eq:mainformula}.
As reviewed in Appendix \ref{sec:coordinates}, the retarded boundary-to-bulk propagator in AdS$_3$ spacetime is given by
\be\label{eq:BBAdS}
G_{B,R}^{\rm AdS}(x,v=0,z;x_1,t_1)  \propto \theta(z-v_1)\textrm{Im}\left\{\frac{z^{\Delta}}{\left[-t_1^2+2 z t_1+(x-x_1)^2+i\epsilon\right]^{\Delta}}\right\}\,,
\ee
and the retarded bulk-to-boundary propagator in BTZ is given by
\beq \label{eq:BBBTZ}
\tilde{G}_{B,R}^{\rm BTZ}( \bar x_2,\bar t_2; \bar x,  \bar v=0,\bar z)\propto \theta(\bar{t}_2-\bar{z})\textrm{Im}
\Bigg\{\frac{\bar z^{\Delta}}{\left[-\bar{t}_2^2+2 \bar z \bar{t}_2+(\bar{x}-\bar{x}_2)^2+i\epsilon\right]^{\Delta}}\Bigg\}\,,
\eeq
in terms of the coordinates
\beq \label{eq:changeccord}
\bar z \equiv z e^{Rx},\quad\bar{x}\equiv \sqrt{R^{-2}-z^2}e^{Rx}\cosh(Rt),\quad \bar{t}\equiv \sqrt{R^{-2}-z^2}e^{Rx}\sinh(Rt),\quad \bar{v}\equiv \bar{t}-\bar{z}\,.
\eeq
The above coordinate transformation maps BTZ locally into AdS$_3$, as reviewed in Appendix \ref{sec:coordinates}.
In the new coordinates, the shell is located on a surface $\bar{v}=0$ where the coordinate relations reduce to $x=\frac{1}{R}\log R\bar{x}$ and $z=\bar z / (R \bar{x})$.

Substituting \eqref{eq:BBAdS} and \eqref{eq:BBBTZ} and performing the change of coordinates \eqref{eq:changeccord},  the integral (\ref{eq:mainformula}) becomes
\beq\label{eq:action}
G_R(x_2,t_2;x_1,t_1)=\int d \bar x\, d\bar{z}\,e^{-S}\,,
\eeq
where
\begin{align}
S= \Delta \Bigg[ &-2\log \bar z+\log \left(R\bar{x}t_1^2-2 t_1 \bar z -R\bar{x}\left(\frac{1}{R}\log\bar{x}-x_1\right)^2\right)  \nonumber \\
&  +\log\left(\frac{1}{R^2}e^{2Rx_2}\sinh^2Rt_2-2 \frac{\bar z}{R} e^{x_2}\sinh Rt_2- \left(\bar{x}-\frac{1}{R}e^{Rx_2}\cosh Rt_2\right)^2\right)\Bigg]
\end{align}
to leading order in $\Delta$.   For simplicity here we consider non-integer $\Delta$ -- the extension to integer $\Delta$ is described in  Appendix~\ref{app:integerdelta}.  The integration contour is determined by the $i\epsilon$ factors in (\ref{eq:BBAdS}) and (\ref{eq:BBBTZ}) and in general it can be rotated to pass through the saddle point
\beq
\partial_{\bar z}S=0,\quad \partial_{\bar{x}}S=0\,.
\eeq
 For the equal space correlation function ($x_1=x_2=0$) the saddle point is at
\begin{align}
\bar{x}&=1/R\,, \\
\bar z &= \frac{2t_1\sinh(\frac{R t_2}{2})}{R t_1\cosh(\frac{Rt_2}{2})+2\sinh(\frac{Rt_2}{2})}\,.
\end{align}
Substituting the saddle point into \eqref{eq:action} gives the retarded correlator
\beq
G_R(0;t_2;0,t_1)\propto e^{-S}\propto \frac{1}{\left[t_1\cosh(\frac{R t_2}{2})-\frac{2}{R}\sinh(\frac{R t_2}{2})\right]^{2 \Delta}},
\eeq
which agrees (up to an undetermined normalization) with the geodesic approximation result \eqref{eq:GRVaidya} for $t_1<0<t_2$.   To make this argument precise one would need to check that the saddle point is on a steepest descent contour, which is nontrivial in the presence of branch cuts.

\setcounter{equation}{0}


\section{Beyond the geodesic approximation}\label{beyond}


In this section we will develop an alternative method for computing propagators in AdS-Vaidya that goes beyond the geodesic approximation  and enables us to compute the result for all spatial momenta ${\bf k}$ and for any conformal dimension $\Delta$.  This method allows us to compute the full spectral function but only gives numerical results.

\subsection{Holographic dictionary without time translation invariance}

Usually the holographic dictionary is discussed in translation invariant backgrounds. The Green functions
then depend only on  the difference $x-x'$ of
two points. When translation invariance
is broken in at least one direction, the Green functions depend explicitly on two points. A standard holographic method
to compute the retarded Green function in the field theory at the boundary
is to study the dual bulk field: choose an infalling boundary
condition at a (coordinate) horizon, solve the field profile from the equation of motion, and then
extract the Green function from the asymptotic behavior.   So we will begin with a brief discussion of
the holographic dictionary for one-point and two-point functions in the case of broken translation
invariance.

Consider a free boson in 3d asymptotically AdS (AAdS) space of action
\beq
S_\phi =-\frac{1}{2}\int d^{3}x\sqrt{-g}\left( \partial_\mu \phi \partial^\mu \phi +m^2\phi^2\right)
+S_{\rm ct}\,,
\eeq
where the counterterm action is given by
\be
S_{\rm ct} = \frac{2 - \Delta}{2}\int dx dt\sqrt{-\gamma}\phi^2|_{z=\epsilon}\,,\label{eq:counter}
\ee
and $\gamma$ is the induced metric on the $z=\epsilon$ surface.\footnote{The counterterm (\ref{eq:counter}) is applicable for the range of scaling dimensions $\Delta<2$, to which we specialize in the following numerical calculations. For larger $\Delta$, and for more complicated bulk field content, one needs to add more counterterms (see e.g. \cite{deHaro:2000xn}).}
In the field theory on the boundary, using standard results of linear response theory applied to AdS/CFT \cite{Kapusta:1989tk, Son, Iqbal:2008by} and large-N factorization, the response of an operator at $(x_2,t_2)$  in the presence of a delta function source at $(x_1, t_1)$ is  given by the retarded correlation function\footnote{Here we are assuming that $\mathcal{O}$ does not have a vacuum expectation value.}
\beq
\langle\mathcal{O}(x_2,t_2)\rangle_{\delta}= i\theta(t_2-t_1)\langle \[\mathcal{O}(x_2,t_2),\mathcal{O}(x_1,t_1)\]\rangle= -G_R(x_2,t_2;x_1,t_1) \, .\label{eq:linearresp}
\eeq
The details of the derivation of (\ref{eq:linearresp}) are included in Appendix \ref{sec:onepoint}. On the other hand, according to the standard holographic dictionary, the one-point function in the boundary theory
is given by
\beq \label{eq:OAdSCFT}
\langle\mathcal{O}(x,t)\rangle_J=\frac{\delta S_{\phi}^{\rm on-shell}}{\delta J(x,t)}=-2\nu \phi_+(x,t)\,,
\eeq
where $J$ and $\phi_+$ denote respectively the leading and subleading asymptotic behavior of the bulk scalar field $\phi$ as $z \to 0$ 
\beq\label{asybehavior}
\phi(x,t,z)=J(x,t)z^{\Delta_-}+\phi_+(x,t)z^{\Delta}+...
\eeq
where $\Delta=1+\nu$, $\Delta_-=\Delta-2\nu$ and $\nu=\sqrt{1+m^2}$.

Comparing \eqref{eq:linearresp} and \eqref{eq:OAdSCFT}, we obtain the two point function
\beq
G_R(x_2,t_2;x_1,t_1)=2\nu\phi_{+,\delta}(x_2,t_2)\,,\label{eq:retardedfrommode}
\eeq
where $\phi_{+,\delta}$ is the subleading mode of a bulk solution that satisfies the boundary condition
$J(x_2,t_2)=\delta(x_2-x_1)\delta(t_2 - t_1)$ for the non-normalizable mode.  A similar formula for the retarded correlator in momentum space
in a time translationally invariant state was obtained in \cite{Iqbal:2008by}. For more discussion on the relationship between the one point function procedure and
the more conventional prescription for obtaining retarded correlators \cite{Son}, see \cite{Iqbal:2008by}. To construct such a solution, it is more convenient
to specify the boundary data and reconstruct the full bulk field profile using the retarded boundary-to-bulk propagator
\beq
\phi (x_2,t_2,z) = \int dx \, dt \, G_{B,R}^{\rm AAdS} (x_2,t_2,z ;x,t) J(x,t) = G_{B,R}^{\rm AAdS}(x_2,t_2,z; x_1,t_1) \, ,\label{eq:bulkfield1}
\eeq
rather than specifying the boundary condition at the Poincar\'e or event horizon and integrating towards the boundary.

In the case of AdS$_3$ spacetime, the boundary-to-bulk propagator is given by
\beq
G_{R,B}^{\rm AdS}(x_2,t_2,z;x_1,t_1)=C_{\Delta}\theta(t_2-t_1)\textrm{Im}\Big(\frac{z}{z^2+(x_2-x_1)^2-(t_2-t_1)^2+i\epsilon}\Big)^{\Delta},\label{eq:adsbbpropag}
\eeq
where $C_{\Delta}=\Gamma(\Delta)/(\pi\Gamma(\Delta-1)$. Using (\ref{eq:bulkfield1}) we obtain the solution $\phi$, whose normalizable mode can be read off, which after using (\ref{eq:retardedfrommode}), leads to the CFT retarded two point function
\beq
G_R^{\rm vacuum}(x_2,t_2;x_1,t_1)=\theta(t_2-t_1)\frac{2(\Delta-1)\Gamma(\Delta)}{\pi\Gamma(\Delta-1)}\textrm{Im}\Big(\frac{1}{(x_2-x_1)^2-(t_2-t_1)^2+i\epsilon}\Big)^{\Delta},
\eeq
This also gives us the two point function in the CFT state dual to the Vaidya spacetime before the shell for $t_1<0$ and $t_2<0$.

In the next section we apply this method to the Vaidya background, the only difference with respect to the pure AdS setup being that to construct the bulk field profile one needs to continue the solution across the null shell. 


\subsection{Numerical construction of the CFT retarded two-point function}

As discussed above, to obtain the retarded boundary correlator, we need to solve the scalar equation of motion $(\Box-m^2)\phi=0$ in the Vaidya metric background. We parametrize the geometry using ingoing Eddington-Finkelstein (EF) coordinates and the metric \eqref{eq:Vaidyaz} and define
\be
h(z,v)=1-\theta(v)R^2 z^2\,.
\ee

Since the metric is translation invariant in the spatial coordinate direction $x$, it is convenient
to perform a Fourier transform. The equation of motion for the scalar field then becomes
\beq
h\partial_z^2\phi+\left(\frac{1}{z}\partial_v\phi-2\partial_v\partial_z\phi\right)+\left(\partial_z h-\frac{h}{z}\right)\partial_z\phi
-\left(\frac{\Delta_-(\Delta_--2)}{z^2}+k^2\right)\phi=0\,.\label{eq:scalareq}
\eeq
This equation is first order in the EF time $v$, and the horizon is a characteristic surface. This means that a solution to the equations of
motion is uniquely specified by a single initial condition at a constant $v$ slice, together with a boundary condition at the AdS boundary.
Also, this means that the only condition one has to impose at the position of the infalling shell $v=0$ is the continuity of the field $\phi$. This makes the collapsing null
shell simpler to analyze than the case of timelike collapse, where one needs to specify two independent junction conditions at the (time-dependent) position of the shell.

Due to its causal nature, for $t_1>0$ the retarded Green function is automatically thermal. Here we focus on correlation functions in the boundary theory with the initial point $(x_1,t_1)$ in the zero temperature (AdS vacuum) region ($t_1<0$).  Note that it is simple to start with the initial condition in the vacuum region: the bulk scalar field solution before the collapse is given by the standard retarded AdS vacuum boundary-to-bulk propagator. It
solves the equation of motion (\ref{eq:scalareq}), it approaches a delta function at the boundary and vanishes outside the future lightcone of $(x_1,t_1)$.

The Fourier transform to mixed $(k,v,z)$ space of the retarded AdS$_3$ boundary-to-bulk propagator is (see Appendix~\ref{sec:retarded})
\begin{align}\label{eq:bbpropag}
G_{B,R}^{\rm AdS}&(k,v,z; t_1)  = \phi(k, v, z) \nonumber \\
&=- C \frac{ \theta(v-t_1) \,z^{\Delta}}{\left[ (v-t_1)^2+2(v-t_1)z\right]^{\frac{2\nu+1}{4}}} |k|^{\nu+\frac 1 2}J_{-\nu-\frac 1 2}\left(|k|\sqrt{(v-t_1)^2+2 (v-t_1)z}\right) ,
\end{align}
where
\be
C = \frac{2^{\frac 1 2 -\nu} \sqrt \pi}{\Gamma[\nu]}\,.
\ee
In particular, at the shell $v=0$ the field profile is given by
\beq \label{eq:initialdata}
\phi(k, v=0, z) = - C \frac{ \theta(-t_1) \, z^{\Delta}}{\left[t_1^2- 2 t_1 z \right]^{\frac{2\nu+1}{4}}} |k|^{\nu+\frac 1 2} J_{-\nu-\frac 1 2}\left(|k|\sqrt{t_1^2-2 t_1 z}\right),
\eeq
and \eqref{eq:bbpropag} indeed satisfies the asymptotic behavior (\ref{asybehavior}) with $J(t)=\delta (t-t_1)$:
\beq
\phi (k,x,v) = \delta(t-t_1) z^{\Delta_-} +\cdots \,,
\eeq
where we recall $v$ reduces to $t$ on the AdS boundary.

In order to construct the full field profile everywhere, we must extend it into the region after the null shell collapse ($v>0$).
As we know the behavior of the field at $v\leq 0$ analytically, the calculation reduces to solving the first order equation
of motion (\ref{eq:scalareq}) in the BTZ background $v>0$, with initial data specified by (\ref{eq:initialdata}). In the following we solve (\ref{eq:scalareq}) numerically. For convenience we choose the radius of the event horizon to be $R=1$ by a conformal rescaling.

For $v>0$, the scalar field equation (\ref{eq:scalareq}) can be solved noting that it can be written as
\beq
\partial_z \left( \frac{\partial_v\phi }{\sqrt z} \right)=\frac{1}{2 \sqrt z}\left[h \partial_z^2\phi+\left(\partial_z h-\frac{h}{z}\right)\partial_z\phi -\left(\frac{\Delta_-(\Delta_--2)}{z^2}+k^2 \right)\phi\right],
\eeq
which can be integrated with respect to $z$ to give
\begin{align} \label{eq:EOMintegrated}
\left. \frac{\partial_v\phi}{\sqrt z} \right|_{z}= \left. \frac{\partial_v\phi}{\sqrt z} \right|_{z_0}+ \int_{z_0}^{z}d\hat{z} \, \frac{1}{2 \sqrt{\hat z}} \left[ h\partial_{\hat{z}}^2\phi +\left(\partial_{\hat{z}} h-\frac{h}{\hat{z}}\right)\partial_{\hat{z}}\phi -\left(\frac{\Delta_-(\Delta_--2)}{\hat{z}^2}+k^2\right)\phi\right]
\end{align}
in terms of the regulated AdS boundary $z =z_0$.  For the numerical results presented in Section~\ref{sect:numerical}, we have set $z_0 = 10^{-10}$.

For computational convenience we define a rescaled field $P(k,v,z)$ as
\beq
\phi(k,v,z)=z^{\Delta_-}P(k,z,v)\,,
\eeq
in terms of which \eqref{eq:EOMintegrated} becomes
\begin{align}\label{eq:integral}
\left.  \partial_v P  \right|_{z} =&\left(\frac{z_0}{z}\right)^{\frac{2\Delta_--1}{2}} \left. \partial_v P  \right|_{z_0} \nonumber \\
&+z^{\frac{1-2\Delta_-}{2}}\int_{z_0}^{z}d\hat{z} \, \frac{\hat{z}^{\frac{2\Delta_--1}{2}}}{2}\left[h \partial_{\hat{z}}^2 P  +\left(\partial_{\hat{z}} h+\frac{(2\Delta-1)h}{\hat{z}}\right)\partial_{\hat{z}} P-\left(\Delta_-^2 +k^2\right) P \right].
\end{align}
To solve this equation numerically, we discretize both space and time
\beq
(v,z)\rightarrow (v[i],z[j])=(i h_v, j h_z)\,,
\eeq
where $(h_v,h_z)=(T/N_v,1/N_z)$, $N_v, N_z$ are the numbers of lattice sites and $T$ is the range of the time coordinate. The field $P(v,z)$ is replaced by discrete  variables
\beq
P(v,z)\rightarrow P\[i,j\]\,,
\eeq
derivatives are replaced by discrete derivatives
\beq
\partial_z P (v,z )\rightarrow \frac{R\[i+1,j\]-R\[i-1,j\]}{2 h_z}\,,
\eeq
and the integral by a sum
\beq
\int dz\rightarrow \sum_i h_z\,.
\eeq
By knowing the value of the field $P$ at some initial constant $v$ slice, we can use the discretized version of (\ref{eq:integral}) to obtain the value of the field at the next time step. Iterating this procedure we obtain the time evolution of the scalar field. 
Alternatively, a similar discretization and integration procedure is efficiently implemented by the Mathematica command NDSolve. The results of the two computations agree and the plots we present in the next Section have been produced using Mathematica.


\subsection{Numerical results}\label{sect:numerical}


\subsubsection{Retarded two-point function at fixed momentum}\label{sect:GRt}

\begin{figure}[h]
\begin{center}
\includegraphics[scale=1.3]{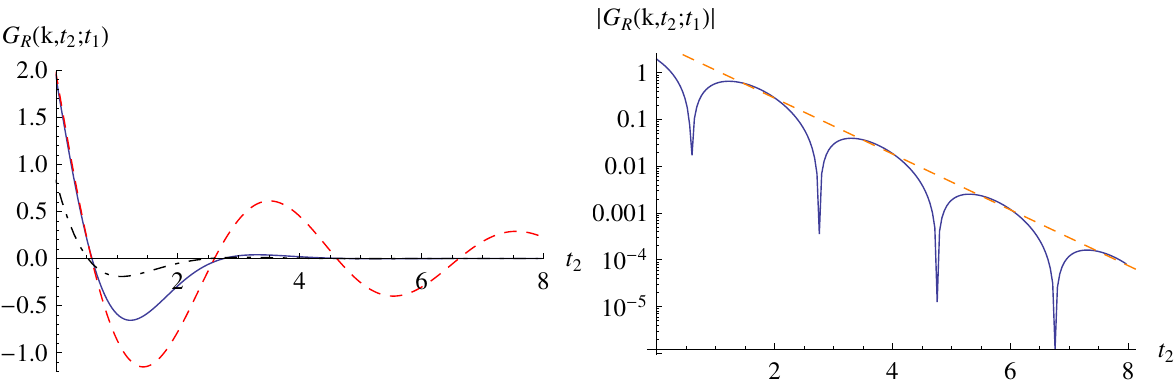}
\caption{\label{fig:corr1} The left panel shows the correlator in the thermalizing background (solid blue curve), together with the vacuum (dashed red) and thermal correlator (dot-dashed black). The right panel compares logarithmic plots of the thermalizing correlator (solid blue) and of the exponential decay of perturbations close to equilibrium set by the imaginary part of the lowest quasinormal mode (orange dot-dashed). In both figures we have fixed $\Delta =5/3$, $t_1=- 2$  and $k = \pi/2$.
}
\end{center}
\end{figure}

The left panel of Figure~\ref{fig:corr1} shows the time evolution of the retarded two-point function for fixed momenta (in blue continuous), together with the time evolution of the vacuum (dashed red) and thermal (dot-dashed black) correlators.
The two-point function in the thermalizing CFT decays faster in time than the vacuum correlator.

A logarithmic plot reveals that the absolute value of the correlator oscillates, with the enveloping curve decaying exponentially in time (right panel of Fig.~\ref{fig:corr1}).
Ingoing field fluctuations in a black hole background are well known to have a quasinormal behavior with a set of discrete oscillation frequencies with negative imaginary parts. These modes have been identified in the AdS/CFT correspondence
as determining the approach to thermal equilibrium in the level of linear response \cite{Horowitz:1999jd}.
Even though in our case we are considering dynamics far from equilibrium, it seems reasonable that at least at large times
the dynamics of the field is dictated by the lowest quasinormal modes. The right hand side of Figure~\ref{fig:corr1} compares the exponential
decay of the field modes with those of the lowest quasinormal mode, which in the black brane background is known to be given by \cite{Son}
\beq
\omega_{quasi}=k-i\Delta.
\eeq
 Also the oscillation frequency of the correlator is given by the real part of the lowest quasinormal mode, here $k$.


\subsubsection{The momentum dependence of the retarded autocorrelator}\label{sect:GRk}

\begin{figure}[h]
\begin{center}
\includegraphics[scale=1.1]{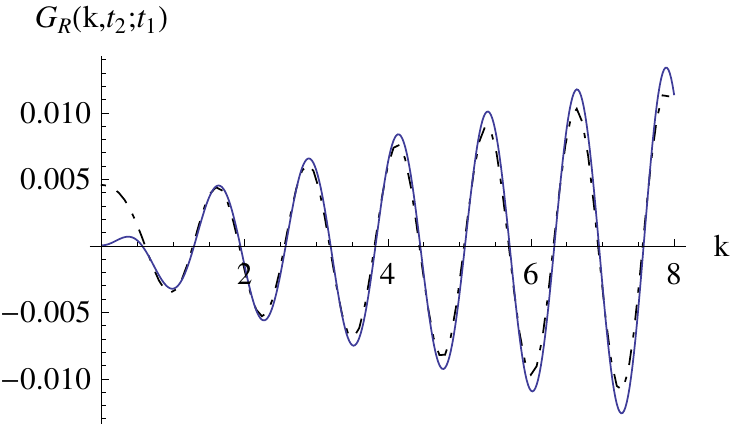}
\caption{\label{fig:fixedt} Numerical result for the correlation function at a fixed time $t_2=5$ and $t_1=-2$ in the thermalizing (solid blue) and thermal (dot-dashed black) background, as a function of momentum. The vacuum curve is not shown in the plot because of much larger amplitude. In the figure we have chosen $\Delta=5/3$.}
\end{center}
\end{figure}

Numerical results for the thermalizing retarded correlation function as a function of momentum, at fixed time, are shown in Figure~\ref{fig:fixedt}.
The plot reveals a modulation by the power law $k^{\nu}$ at large momentum which is also present in the conformal field theory vacuum 
\beq
G_R^{\rm vacuum}(k,t_2 ; t_1)\propto -\theta(t_2 -t_1 )\left(\frac{|k|}{t_2 -t_1}\right)^{\nu+\frac{1}{2}}J_{-\nu-\frac{1}{2}}(|k|(t_2 -t_1))\,,
\eeq
which can be obtained by Fourier transforming \eqref{eq:GRvacuum}.

For small momenta the thermalizing retarded correlator in Figure~\ref{fig:fixedt} differs significantly from the thermal correlator. In particular, the thermal 
correlator is a non-vanishing function of $k$ at small $k$, while the thermalizing correlator vanishes at small $k$. This non thermal form at small momenta is in qualitative agreement with the "top-down" thermalization pattern obtained from many of the holographic thermalization studies.


\subsubsection{Spectral function}\label{sect:rhoT}

\begin{figure}[h]
\begin{center}
\includegraphics[scale=1.3]{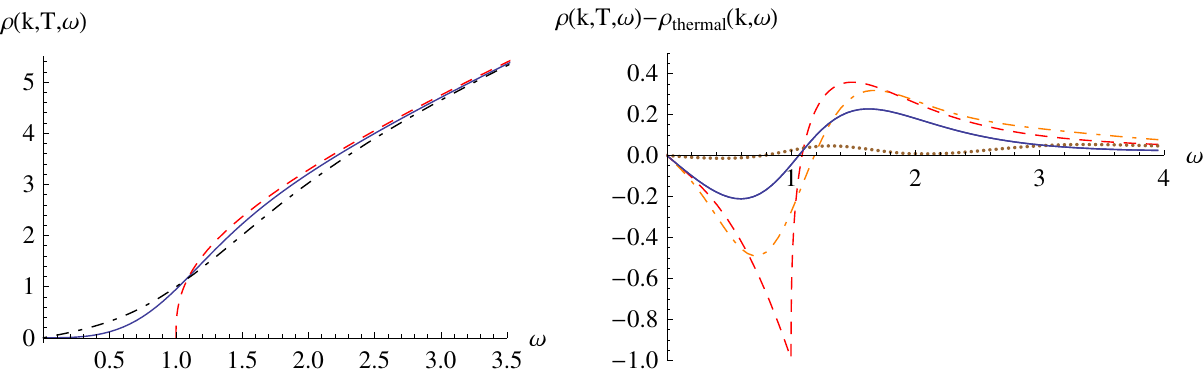}
\caption{\label{fig:wavespectral}  Wigner transformed correlation function without (left) or with (right) the thermal two-point function subtracted. The different curves correspond to different values of the mean time $T$: the vacuum $T=-\infty$ (red, dashed), $T=-1$ (orange, dot-dashed), $T=0$ (blue, solid), $T=1$ (brown, dotted), the thermal $T = \infty$ (black, dot-dashed). In the figure we have set $k=1$ and $\Delta=11/8$.  Note that in empty AdS the spectral function vanishes for $\omega < k$ as it should.}
\end{center}
\end{figure}

As in Sections~\ref{time} and \ref{sect:spectralgeodesic}, we introduce a mean time $T=t_1+t_2$ and a time separation $t=t_2-t_1$.
Fourier transforming $G_R(k, T, \omega)$ with respect to $t$
\beq
G_R(k, T, \omega)=\int_{-\infty}^{\infty} \text{d}t\,e^{i\omega t} G_R(k, t_2; t_1)\,,
\eeq
we obtain the time-dependent spectral function
\beq
\rho(k, T,\omega)=-2\,\textrm{Im}G_R(k, T, \omega)\,.
\eeq
In Figure~\ref{fig:wavespectral}, we display the frequency dependence of the spectral function for fixed momentum and various average times $T$, obtained by Wigner transforming our numerical retarded correlator. Unlike in the harmonic oscillator example, the spectral function here is a smooth function with no quasiparticle peaks. In the vacuum the spectral function has the simple form (\ref{eq:vacuumthermalrho}): $\rho\propto \theta(\omega-k)(\omega^2-k^2)^{\nu}$. The momentum $k$ determines the frequency when the spectral function obtains support. In the thermal state, the spectral function has support all the way down to $\omega=0$ and at large $\omega$ approaches the vacuum form. The time dependent spectral function in Fig.~\ref{fig:wavespectral} smoothly interpolates between the vacuum and thermal forms.


\section{Conclusions and outlook}
\label{conclusions}

We have studied some aspects of thermalization following a holographic quench in the strongly coupled quantum field theory dual to the AdS$_3$ geometry. In this scenario, energy is suddenly injected in the ultraviolet in the form of a thin shell of null energy that falls in the bulk AdS$_3$ space eventually forming a BTZ black hole. In particular, we have investigated unequal-time two-point functions of the boundary field theory via their holographic duals, either in the semiclassical approximation by calculating the length of the bulk geodesic connecting the endpoints of the boundary correlator, or by solving the appropriate wave equation in the time-dependent bulk geometry.

One quantity of particular interest is the spectral function of the correlator, which changes from the vacuum spectral function to the thermal spectral function during thermalization. While causality dictates that the vacuum spectral function $\rho(\omega,k)$ exhibits a gap $\omega_{\rm min}=k$ caused by the absence of space-like excitations, the thermal spectral function exhibits no gap due to the presence of real excitations in the thermal state. In the case of the quench followed by thermalization, the time-dependent spectral function smoothly changes from the vacuum spectral function to the thermal one. Because it involves a Fourier transform covering an infinite time range, the time-dependent spectral function deviates from the vacuum spectral function even before the quench, especially for low frequencies, where it does not exhibit a gap.

The techniques developed in this article enable the calculation of many quantities of interest in the thermalization process. For example, one could study the time dependence of the occupation number of field modes in the boundary gauge theory.

\setcounter{equation}{0}


\section*{Acknowledgments}

We would like to thank Jan de Boer, Federico Galli, Andreas Sch\"afer, Masaki Shigemori and Wieland Staessens for discussions on related issues.

This research is supported by DOE grants  DE-FG02-05ER41367 and DE-FG02-95ER40893, by the Belgian Federal Science Policy Office through the Interuniversity Attraction Poles IAP VI/11 and P7/37, by FWO-Vlaanderen through projects G.0651.11 and G011410N, by the Academy of Finland grant 1127482,  by the National Science Foundation under Grant No.\ PHY11-25915, by grants from the Icelandic Research Fund and the University of Iceland Research Fund, and by the European Science Foundation Holograv Network. 

AB is a Postdoctoral Researcher of FWO-Vlaanderen. JV is Aspirant FWO. VB, AB, BC and EKV are grateful to the Lorentz Center; VB and BC to the KITP; BC to the University of Amsterdam; BC and EKV to the Newton Institute for hospitality and partial support while this work was in progress.

\setcounter{equation}{0}


\appendix


\section{A notion of time-dependent entropy}\label{app:entropy}

Here we summarize the arguments in \cite{BPM} that lead to the entropy formula given in Sec.~\ref{spectral}.      Let us consider particles with Bose statistics indexed by a macroscopic quantum number $I$.   Also suppose that the states index by $I$ have a degeneracy $g_I$, and that there are $n_I$ particles with quantum number $I$.  In both quantum physics and in statistical physics we would really have a probability distribution over $n_I$, but we will assume that $n_I \gg 1$ and the  fluctuations are much smaller than the mean so that we can ignore them.  Thus we will take $n_I$ as given and will not integrate over the distribution of this variable.

Then, the number of ways of distributing the $n_I$ particles over the $g_I$ microstates is
\be
W(n_I) = {(n_I + g_I - 1)!  \over n_{I}! \, (g_I -1)!} \, .
\ee
The numerator counts the number of ways of putting $g_I - 1$ partitions in a set of $n_I$ particles and the denominator comes from the indistinguishability of the particles.  Here we are assuming that the particles are bosons.   Then the number of total possible states of the system is
\be
\Gamma(\{ n_I \}) = \prod_I W(n_I).
\ee
Now let $\alpha$ be a particular micro state with macro occupation numbers $\{ n_I \}$.   Assuming that all the  microstates are equally probable,
\be
P_{\{ n_I \}}(\alpha) =   {1 \over \Gamma(\{ n_I \})}
\ee
is the probability of the microstate $\alpha$

Thus we can write a formula for the von Neumann entropy
\be
S = - \sum_\alpha P_{\{ n_I \}}(\alpha) \ln P_{\{ n_I \}}(\alpha) = \ln \Gamma(\{ n_I \}) = \sum_I \ln W(n_I)
\ee
where we used the fact that $\sum_{\alpha} P_{\{ n_I \}}(\alpha)  = 1$.    Assuming that $n_I, g_I \gg 1$ we can use Stirling's formula to write
\be
S = \sum_I (n_I + g_I ) \ln (n_I + g_I )  - g_I  \ln g_I  - n_I  \ln n_I ,
\ee
where we also dropped the $-1$ in $n_I + g_I - 1$ and $g_I - 1$.      We can simplify further by factoring out $g_I$:
\be
S = \sum_I g_I \left[ \left({n_I \over g_I }+1\right) \ln \left( g_I \left({n_I \over g_I}+ 1 \right) \right)
- \ln (g_I) - \left({n_I \over g_I } \right) \ln \left( g_I \left({n_I \over g_I} \right)\right)
\right]
\ee
There are some cancellations, and then we can write
\be
S = \sum_I g_I \left[ \left({n_I \over g_I} + 1 \right) \ln \left({n_I \over g_I} + 1 \right) -
\left({n_I \over g_I}  \right) \ln \left({n_I \over g_I} \right)
\right].
\ee
Here $n_I/g_I$ is the average occupation number of microstates.

To make contact with the notions introduced in section~\ref{spectral}, we interpret $g_I = \rho({\bf k},\omega,T)$, i.e., we take $g_I$ to be the time dependent density of states with macroscopic quantum numbers ${\bf k},\omega$.    We also take $n_I = n({\bf k},\omega,T) \rho({\bf k},\omega,T)$, i.e. the total occupation number of modes with ${\bf k},\omega$.   We then get
\be
S(T) = \int d\omega \, d{\bf k} \, \rho({\bf k},\omega,T) \, s({\bf k},\omega,T)
\ee
where
\be
s({\bf k},\omega,T) =
\left( n({\bf k},\omega,T) + 1\right)
\ln \left( n({\bf k},\omega,T) + 1\right)
-
\left( n({\bf k},\omega,T)  \right)
\ln \left( n({\bf k},\omega,T)\right).
\ee

To illustrate this formula, consider the simple case when the spectral density is a sum of (on-shell) delta functions:
\be
\rho({\bf k},\omega,T) = \sum_i \delta(\omega-\omega_i({\bf k})) ,
\ee
where the sum may account for any degeneracy by having several $\omega_i({\bf k})$ have the same value. Then in equilibrium:
\be
n({\bf k},\omega) = n_B(\omega) \rho({\bf k},\omega) = \sum_i n_B(\omega_i({\bf k})) \delta(\omega-\omega_i({\bf k})) ,
\ee
where $n_B(\omega) = (e^{\beta\omega}-1)^{-1}$ is the Bose distribution. Thus $n_I =n_B(\omega)$ in equilibrium only depends on $\omega$, not on ${\bf k}$, and it is independent of the level density and the degeneracy of levels. The formula for the entropy then becomes:
\be
s(\omega) = (n_B(\omega)+1)\ln(n_B(\omega)+1) - n_B(\omega) \ln n_B(\omega) .
\ee
The quantity $s$ accounts for the entropy associated with the different possible occupation numbers of one bosonic mode of energy $\omega$.

\setcounter{equation}{0}


\section{Euclidean AdS$_3$ and BTZ}\label{app:geodesics}


\subsection{Geodesic length in Euclidean AdS$_{3}$ and BTZ}

We here explicitly derive the renormalised length of a geodesic between two boundary points in Euclidean BTZ and AdS$_{3}$.  This serves both to introduce the Wick rotation in this setting, as well as to derive a number of expressions that will be used in the main text.



We parametrize Euclidean BTZ using Poincar\'e coordinates in which the metric reads
\begin{equation}\label{eq:EBTZmetric}
ds^{2}=(r^{2}-R^{2})dy^{2} + \frac{dr^{2}}{r^{2}-R^{2}}+r^{2}dx^{2},
\end{equation}
where $R<r$, the surface $r=R$ is the black hole horizon and $r \to \infty$ represents the spacetime boundary. Since Euclidean AdS$_3$ is recovered from \eqref{eq:EBTZmetric} by sending $R \to 0$, we will work out explicitly the BTZ case alone and take the AdS limit only at the end.

The Euclidean BTZ geodesics are given by
\begin{eqnarray}
r(\lambda)^{2}&=&\Gamma_{-}^{2}+(\Gamma_{+}^{2}-\Gamma_{-}^{2})\cosh^{2}(\lambda-\lambda_{0})\,, \\
x(\lambda)&=&x_{0}+\frac{1}{R}\text{arctanh}\left(\frac{\Gamma_{-}}{\Gamma_{+}}\tanh(\lambda-\lambda_{0})\right), \\
y(\lambda)&=&y_{0}+\frac{1}{R}\text{arctan}\left(\sqrt{\frac{R^{2}-\Gamma_{-}^{2}}{\Gamma_{+}^{2}-R^{2}}}\tanh(\lambda-\lambda_{0})\right)\,,
\end{eqnarray}
for constants $0<\Gamma_{-}^{2}<R^{2}<\Gamma_{+}^{2}$ and $x_0$, $y_0$, $\lambda_0$.
As a function of $r$, we find the following branches ($r\in[\Gamma_{+},\infty[$),
\begin{eqnarray}
x_{\pm}(r)&=&x_{0}\pm\frac{1}{R}\text{arctanh}\left(\frac{\Gamma_{-}}{\Gamma_{+}}\sqrt{\frac{r^{2}-\Gamma_{+}^{2}}{r^{2}-\Gamma_{-}^{2}}}\right), \label{eq:xpm} \\
y_{\pm}(r)&=&y_{0}\pm\frac{1}{R}\text{arctan}\left(\sqrt{\frac{R^{2}-\Gamma_{-}^{2}}{\Gamma_{+}^{2}-R^{2}}}\sqrt{\frac{r^{2}-\Gamma_{+}^{2}}{r^{2}-\Gamma_{-}^{2}}}\right), \label{eq:ypm}\\
\lambda_{\pm}(r)&=&
\lambda_{0}\pm\ln\left(\frac{\sqrt{r^{2}-\Gamma_{-}^{2}}+\sqrt{r^{2}-\Gamma_{+}^{2}}}{\sqrt{\Gamma_{+}^{2}-\Gamma_{-}^{2}}}\right).\label{eq:lambdapm}
\end{eqnarray}
The separation between two boundary insertion points $(x_{1},y_{1},r=\infty)$ and $(x_{2},y_{2},r=\infty)$ is given by
\begin{eqnarray}
\Delta x&=&x(\lambda\rightarrow\infty)-x(\lambda\rightarrow-\infty)=\frac{2}{R}\text{arctanh}\left(\frac{\Gamma_{-}}{\Gamma_{+}}\right), \\
\Delta y&=&y(\lambda\rightarrow\infty)-y(\lambda\rightarrow-\infty)=\frac{2}{R}\text{arctan}\left(\sqrt{\frac{R^{2}-\Gamma_{-}^{2}}{\Gamma_{+}^{2}-R^{2}}}\right),
\end{eqnarray}
and the length of the geodesic can be found to be
\begin{eqnarray}
\Delta\lambda&=&\lambda_{+}(r\rightarrow r_{0})-\lambda_{-}(r\rightarrow r_{0})= 2\ln\left(\frac{\sqrt{r_{0}^{2}-\Gamma_{-}^{2}}+\sqrt{r_{0}^{2}-\Gamma_{+}^{2}}}{\sqrt{\Gamma_{+}^{2}-\Gamma_{-}^{2}}}\right) 
\end{eqnarray}
in terms of a regularized AdS boundary at $r =r_0$.
In the limit $r_{0}\rightarrow\infty$ this reduces to
\begin{equation} \label{eq:intermediate}
\Delta\lambda=2\ln\left(\frac{2r_{0}}{\sqrt{\Gamma_{+}^{2}-\Gamma_{-}^{2}}}\right).
\end{equation}

We now pause for a moment and consider the $R \to 0$ AdS limit. In this limit, the geodesics branches \eqref{eq:xpm}-\eqref{eq:lambdapm} become
\begin{eqnarray}
x_{\pm}(r)&=&x_{0}\pm\frac{1}{\Sigma}\cos(\sigma)\sqrt{1-\frac{\Sigma^{2}}{r^{2}}}, \\
y_{\pm}(r)&=&y_{0}\pm\frac{1}{\Sigma}\sin(\sigma)\sqrt{1-\frac{\Sigma^{2}}{r^{2}}}, \\
\lambda_{\pm}(r)&=&\lambda_{0}\pm\text{arccosh}\left(\frac{r}{\Sigma}\right)=\lambda_{0}\pm\ln\left(\frac{r+\sqrt{r^{2}-\Sigma^{2}}}{\Sigma}\right).
\end{eqnarray}
where $r\in[\Sigma,\infty[$ and
\begin{equation}
\lim_{R\rightarrow0}\Gamma_{+}^{2}\equiv \Sigma^{2}
\qquad\text{and}\qquad
\lim_{R\rightarrow0}\frac{\Gamma_{-}^{2}}{R^{2}} \equiv \cos^{2}(\sigma)\,.
\end{equation}
The geodesic length \eqref{eq:intermediate} thus reduces for $R\to 0$ to
\begin{equation} \label{eq:AdSrenorm}
\Delta\lambda=2\ln\left(\frac{2r_{0}}{\Sigma}\right)=2\ln\left(r_{0}\sqrt{(\Delta x)^{2}+(\Delta y)^{2}}\right).
\end{equation}
We renormalize the geodesic length by subtracting the divergent contribution $2  \ln r_0$, so that the renormalised length of the geodesic in Euclidean AdS is given by
\begin{equation}\label{eq:AdSlength}
\delta\mathcal{L}_{\rm EAdS} \equiv \ln\left[(\Delta x)^{2}+(\Delta y)^{2}\right].
\end{equation}

In a similar way, we renormalize the Euclidean BTZ result \eqref{eq:intermediate} and obtain
\begin{equation}\label{eq:BTZlength}
\delta\mathcal{L}_{\rm EBTZ} \equiv \Delta\lambda-2\ln(r_{0})=\ln\left(\frac{4}{\Gamma_{+}^{2}-\Gamma_{-}^{2}}\right)=\ln\left[\frac{4}{R^{2}}\left(\sinh^{2}\left(\frac{R\Delta x}{2}\right)+\sin^{2}\left(\frac{R\Delta y}{2}\right)\right)\right].
\end{equation}


\subsection{Two point functions of vacuum and thermal CFT}

The Euclidean two-point functions can be computed in the geodesic approximation from the renormalised geodesic lengths that were obtained in the previous section.
A suitable Wick rotation leads to the time-ordered and retarded two-point functions in Lorentzian signature. Below we also briefly comment on how to extract the spectral function and occupation number from these results.


\subsubsection{Euclidean two point functions}

Remember that, in the geodesic approximation, the renormalised two-point function in the dual CFT equals
\begin{equation}
\langle\mathcal{O}(x_{1},y_{1})\mathcal{O}(x_{2},y_{2})\rangle_{\text{ren}}\sim e^{-\Delta \delta\mathcal{L} }\,.
\end{equation}
From \eqref{eq:AdSlength} and \eqref{eq:BTZlength}, we then find the (Euclidean) vacuum two-point function
\begin{equation}
G_E^{\text{vacuum}}(x_{2},y_{2};x_{1},y_{1})=\frac{1}{\left[(\Delta x)^{2}+(\Delta y)^{2}\right]^{\Delta}},
\end{equation}
and the (Euclidean) thermal two point function
\begin{equation}
G_E^{\text{thermal}}(x_{2},y_{2};x_{1},y_{1})=\frac{1}{\left[\frac{4}{R^{2}}\left(\sinh^{2}\left(\frac{R\Delta x}{2}\right)+\sin^{2}\left(\frac{R\Delta y}{2}\right)\right)\right]^{\Delta}} \,.
\end{equation}
where the first can be obtained from the second in the $R\to0$ limit of vanishing black hole mass.
Despite the fact that we have used a geodesic approximation, the result for the vacuum and thermal two-point function is exact, being fully constrained by conformal invariance (apart from an overall scaling).


\subsubsection{Lorentzian two point functions}

To go from Euclidean coordinates $(x,y,r)$ to Lorentzian coordinates $(x,t,r)$, we perform the Wick rotation $y=it$. It is well known that by Wick rotating the Euclidean two point function, one finds the time-ordered (Feynman) two point function
\begin{equation}
iG_F(x_{2},t_{2};x_{1},t_{1})=G_E(x_{2},it_{2};x_{1},it_{1})\,.
\end{equation}
For non integer scaling dimension $\Delta$, this leads to\footnote{For the case of integer $\Delta$, see Appendix~\ref{app:integerdelta}.}
\begin{eqnarray}
iG_F^{\text{vacuum}}(x_{2},t_{2};x_{1},t_{1})&=&\frac{\theta\left[(\Delta x)^{2}-(\Delta t)^{2}\right]}{\left[(\Delta x)^{2}-(\Delta t)^{2}\right]^{\Delta}}+\frac{\theta\left[(\Delta t)^{2}-(\Delta x)^{2}\right]}{\left[(\Delta t)^{2}-(\Delta x)^{2}\right]^{\Delta}}e^{-i\pi\Delta}, \label{eq:GFvacuum}\\
iG_F^{\text{thermal}}(x_{2},t_{2};x_{1},t_{1})&=&\frac{\theta\left[(\Delta x)^{2}-(\Delta t)^{2}\right]}{\left[\frac{4}{R^{2}}\left(\sinh^{2}\left(\frac{R\Delta x}{2}\right)-\sinh^{2}\left(\frac{R\Delta t}{2}\right)\right)\right]^{\Delta}} \nonumber \\
&&\qquad\qquad+\frac{\theta\left[(\Delta t)^{2}-(\Delta x)^{2}\right]}{\left[\frac{4}{R^{2}}\left(\sinh^{2}\left(\frac{R\Delta t}{2}\right)-\sinh^{2}\left(\frac{R\Delta x}{2}\right)\right)\right]^{\Delta}}e^{-i\pi\Delta}\,. \label{eq:GFthermal} \qquad\qquad
\end{eqnarray}
The identity,
\begin{equation}
G_F(x_{2},t_{2};x_{1},t_{1})+(G_F(x_{2},t_{2};x_{1},t_{1}))^{*}= G_R(x_{2},t_{2};x_{1},t_{1})+G_A(x_{2},t_{2};x_{1},t_{1})\,,
\end{equation}
where $G_{R/A}$ denotes the retarded (advanced) two-point function, allows to deduce
\begin{equation}
G_R(x_{2},t_{2};x_{1},t_{1})=\theta(t_{1}-t_{2})[G_F(x_{2},t_{2};x_{1},t_{1})+(G_F(x_{2},t_{2};x_{1},t_{1}))^{*}]\,.
\end{equation}
In our case, this results in
\begin{eqnarray}
G_{R}^{\text{vacuum}}(x_{2},t_{2};x_{1},t_{1})&=&-2\sin\left(\pi\Delta\right)\theta\left(\Delta t\right)\frac{\theta\left[(\Delta t)^{2}-(\Delta x)^{2}\right]}{\left[(\Delta t)^{2}-(\Delta x)^{2}\right]^{\Delta}}, \label{eq:GRvacuum}\\
G_{R}^{\text{thermal}}(x_{2},t_{2};x_{1},t_{1})&=&-2\sin\left(\pi\Delta\right)\theta\left(\Delta t\right)\frac{\theta\left[(\Delta t)^{2}-(\Delta x)^{2}\right]}{\left[\frac{4}{R^{2}}\left(\sinh^{2}\left(\frac{R\Delta t}{2}\right)-\sinh^{2}\left(\frac{R\Delta x}{2}\right)\right)\right]^{\Delta}}\,. \label{eq:GRthermal}\qquad\qquad
\end{eqnarray}


\subsubsection{Spectral function}

Performing a Fourier transform
\begin{equation}
G(k,\omega)=\int_{-\infty}^{+\infty}\text{d}x\int_{-\infty}^{+\infty}\text{d}t\,e^{ikx}e^{i\omega t}G(x,t),
\end{equation}
where we take $x=\Delta x=x_{2}-x_{1}$ and $t=\Delta t=t_{2}-t_{1}$, we can calculate the spectral function $\rho(k,\omega)$ and occupation number $n(k,\omega)$ from the identities \eqref{eq:defrho} and \eqref{eq:defn}
\begin{eqnarray}
\rho(k,\omega)&=&-2\,\text{Im}G_{R}(k,\omega)\,, \\
(1+2n(k,\omega))\rho(k,\omega)&=&-2\,\text{Im}G_F(k,\omega)\,. \label{eq:defn2}
\end{eqnarray}
The Fourier transforms of \eqref{eq:GFvacuum},  \eqref{eq:GRvacuum}, \eqref{eq:GFthermal} and \eqref{eq:GRthermal} read respectively
\begin{eqnarray}
G_F^{\text{vacuum}}(k,\omega)
&=&-\pi\frac{\Gamma(1-\Delta)}{\Gamma(\Delta)}\left[\left(\frac{\omega^{2}-k^{2}}{4}\right)^{\Delta-1}\theta\left(\omega^{2}-k^{2}\right)\left[i\sin(\pi\Delta)-\cos(\pi\Delta)\right]\right. \nonumber \\
&&\qquad\qquad\qquad\qquad\qquad\qquad\quad\left.+\left(\frac{k^{2}-\omega^{2}}{4}\right)^{\Delta-1}\theta\left(k^{2}-\omega^{2}\right)\right], \\
G_R^{\text{vacuum}}(k,\omega)
&=&-\pi\frac{\Gamma(1-\Delta)}{\Gamma(\Delta)}\left[\left(\frac{\omega^{2}-k^{2}}{4}\right)^{\Delta-1}\theta\left(\omega^{2}-k^{2}\right)\left[i\,\text{sign}(\omega)\sin(\pi\Delta)-\cos(\pi\Delta)\right] \right. \nonumber \\
&&\qquad\qquad\qquad\qquad\qquad\qquad\quad\left.+\left(\frac{k^{2}-\omega^{2}}{4}\right)^{\Delta-1}\theta\left(k^{2}-\omega^{2}\right)\right], \\
G_F^{\text{thermal}}(k,\omega)
&=&\frac{R^{2\Delta-2}}{2\sin(\pi\Delta)(\Gamma(\Delta))^{2}}\left|\Gamma\left(\frac{\Delta}{2}+\frac{i(\omega+k)}{2R}\right)\Gamma\left(\frac{\Delta}{2}+\frac{i(\omega-k)}{2R}\right)\right|^{2} \nonumber \\
&&\quad\left\{\cos(\pi\Delta)\cosh\left(\frac{\pi\omega}{R}\right)-\cosh\left(\frac{\pi k}{R}\right)-i\sin(\pi\Delta)\cosh\left(\frac{\pi\omega}{R}\right)\right\}, \\
G_R^{\text{thermal}}(k,\omega)
&=&\frac{R^{2\Delta-2}}{2\sin(\pi\Delta)(\Gamma(\Delta))^{2}}\left|\Gamma\left(\frac{\Delta}{2}+\frac{i(\omega+k)}{2R}\right)\Gamma\left(\frac{\Delta}{2}+\frac{i(\omega-k)}{2R}\right)\right|^{2} \nonumber \\
&&\quad\left\{\cos(\pi\Delta)\cosh\left(\frac{\pi\omega}{R}\right)-\cosh\left(\frac{\pi k}{R}\right)-i\sin(\pi\Delta)\sinh\left(\frac{\pi\omega}{R}\right)\right\},
\end{eqnarray}
and agree with those computed in \cite{Son} up to an overall normalization.

From these expressions, we can deduce that\footnote{For half integer values of $\Delta$, we can use the following expressions:
\begin{equation}
\left|\Gamma\left(1+ix\right)\right|^{2}=\frac{\pi x}{\sinh(\pi x)}
\text{,}\qquad
\left|\Gamma\left(\frac{1}{2}+ix\right)\right|^{2}=\frac{\pi}{\cosh(\pi x)}
\quad\text{and}\quad
\Gamma(x+1)=x\Gamma(x).
\end{equation}}
\begin{eqnarray} \label{eq:vacuumthermalrho}
\rho_{\text{vacuum}}(k,\omega)&=&\frac{2\pi^{2}}{(\Gamma(\Delta))^{2}}\left(\frac{\omega^{2}-k^{2}}{4}\right)^{\Delta-1}\theta\left(\omega^{2}-k^{2}\right)\text{sign}(\omega), \nonumber \\
\rho_{\text{thermal}}(k,\omega)
&=&\frac{R^{2\Delta-2}}{(\Gamma(\Delta))^{2}}\left|\Gamma\left(\frac{\Delta}{2}+\frac{i(\omega+k)}{2R}\right)\Gamma\left(\frac{\Delta}{2}+\frac{i(\omega-k)}{2R}\right)\right|^{2}\sinh\left(\frac{\pi\omega}{R}\right).\qquad\qquad
\end{eqnarray}
Integrating over all momenta gives the results
\begin{eqnarray}
\rho_{\text{vacuum}}(\omega)&=&\frac{4\pi^{2}}{\Gamma(2\Delta)}|\omega|^{2\Delta-1}\text{sign}(\omega), \\
\rho_{\text{thermal}}(\omega)&=&\frac{4\pi R^{2\Delta-1}}{\Gamma(2\Delta)}\left|\Gamma\left(\Delta+\frac{i\omega}{R}\right)\right|^{2}\sinh\left(\frac{\pi\omega}{R}\right)\,,
\end{eqnarray}
while for the occupation number, using \eqref{eq:defn2}, we recover Bose-Einstein distribution
\begin{equation}
n_{\text{thermal}}(k,\omega)=\frac{1}{2}\left(\frac{\text{Im}G_F^{\text{thermal}}(k,\omega)}{\text{Im} G_R^{\text{thermal}}(k,\omega)}-1\right)=\frac{1}{e^{\frac{2\pi\omega}{R}}-1}\,.
\end{equation}

\setcounter{equation}{0}


\section{Notes on the case of integer Delta}\label{app:integerdelta}


\subsection{Euclidean signature}

Consider a conformal field theory in $d$ dimensions. The Euclidean two-point function of a scalar operator with conformal weight $\Delta$ is given by
\begin{equation}
G(\mathbf{x})=\frac{1}{(\mathbf{x}^{2})^{\Delta}}\,.
\end{equation}
The Fourier transform of this function is 
\begin{align}
G(\mathbf{k})&=\int\text{d}^{d}\mathbf{x}\,G(\mathbf{x})e^{i\mathbf{k}\cdot\mathbf{x}}
=\frac{1}{\Gamma\left(\Delta\right)}\int\text{d}^{d}\mathbf{x}\,e^{i\mathbf{k}\cdot\mathbf{x}}\int_{0}^{\infty}\text{d}s\,s^{\Delta-1}e^{-s\mathbf{x}^{2}} \nonumber \\
=& \frac{1}{\Gamma\left(\Delta\right)}\int_{0}^{\infty}\text{d}s\,s^{\Delta-1}\left(\frac{\pi}{s}\right)^{\frac{d}{2}}e^{-\frac{\mathbf{k}^{2}}{4s}} 
=\frac{\pi^{\frac{d}{2}}}{\Gamma\left(\Delta\right)}\int_{0}^{\infty}\text{d}u\,u^{\frac{d}{2}-\Delta-1}e^{-\frac{\mathbf{k}^{2}}{4}u}\nonumber \\
=&\pi^{\frac{d}{2}}\frac{\Gamma\left(\frac{d}{2}-\Delta\right)}{\Gamma\left(\Delta\right)}\left(\frac{\mathbf{k}^{2}}{4}\right)^{\Delta-\frac{d}{2}}\,,
\label{ScalarFourierTransformDeltaLessThandOver2}
\end{align}
where we have used the Gaussian integral
\begin{equation}
\int\text{d}x\,e^{ikx}e^{-\beta x^{2}}=\sqrt{\frac{\pi}{\beta}}e^{-\frac{k^{2}}{4\beta}}\,,
\end{equation}
the relation (valid for $\alpha>0$)
\begin{equation}
\frac{1}{y^{\alpha}}=\frac{1}{\Gamma(\alpha)}\int_{0}^{\infty}\text{d}s\,s^{\alpha-1}e^{-sy}\,,
\end{equation}
and made the substitution $s=1/u$

This derivation is only strictly valid for $d/2>\Delta>0$, which is not of physical interest to us since we always have $\Delta\ge d/2$. However, one can analytically continue the result in $\Delta$ to all positive real values for which $\Delta-d/2$ is not zero or a positive integer. We could also use a regulator $\xi$, such that
\begin{equation}
G_{\xi}(\mathbf{x})=\frac{1}{(\mathbf{x}^{2}+\xi^{2})^{\Delta}}
\end{equation}
and determine the Fourier transform 
\begin{align}
G_{\xi}(\mathbf{k})&=\int\text{d}^{d}\mathbf{x}\,G_{\xi}(\mathbf{x})e^{i\mathbf{k}\cdot\mathbf{x}}
=\frac{1}{\Gamma\left(\Delta\right)}\int\text{d}^{d}\mathbf{x}\,e^{i\mathbf{k}\cdot\mathbf{x}}\int_{0}^{\infty}\text{d}s\,s^{\Delta-1}e^{-s\mathbf{x}^{2}}e^{-s\xi^{2}} \nonumber \\
&=\frac{1}{\Gamma\left(\Delta\right)}\int_{0}^{\infty}\text{d}s\,s^{\Delta-1}\left(\frac{\pi}{s}\right)^{\frac{d}{2}}e^{-\frac{\mathbf{k}^{2}}{4s}}e^{-s\xi^{2}}
=\frac{\pi^{\frac{d}{2}}}{\Gamma\left(\Delta\right)}\int_{0}^{\infty}\frac{\text{d}u}{u^{\Delta-\frac{d}{2}+1}}e^{-\frac{\mathbf{k}^{2}}{4}u}e^{-\frac{\xi^{2}}{u}} \nonumber \\
&=\frac{2\pi^{\frac{d}{2}}}{\Gamma\left(\Delta\right)}\left(\frac{\sqrt{\mathbf{k}^{2}}}{2\xi}\right)^{\Delta-\frac{d}{2}}K_{\Delta-\frac{d}{2}}(\sqrt{\mathbf{k}^{2}}\,\xi)\,,
\end{align}
where we have used the following expression for the modified Bessel function (valid for all $a$ and $b$)
\begin{equation}
2\left(\frac{a}{b}\right)^{\nu}K_{\nu}(2ab)=\int_{0}^{\infty}\frac{\text{d}s}{s^{\nu+1}}\exp\left(-sa^{2}-\frac{b^{2}}{s}\right).
\end{equation}
If $\nu=\Delta-d/2$ is non-integer, then we can use the following series expansion around $x=0$.
\begin{equation}
K_{\nu}(x)=\frac{\pi}{2}\frac{\left(I_{-\nu}(x)-I_{\nu}(x)\right)}{\sin(\nu\pi)}
\qquad\text{and}\qquad
I_{\nu}(x)=\left(\frac{x}{2}\right)^{\nu}\sum_{k=0}^{\infty}\frac{\left(\frac{x}{2}\right)^{2k}}{k!\Gamma(\nu+k+1)}.
\end{equation}
Thus we find that,
\begin{align}
G_{\xi}(\mathbf{k})&=\frac{2\pi^{\frac{d}{2}}}{\Gamma\left(\Delta\right)}\left(\frac{\sqrt{\mathbf{k}^{2}}}{2\xi}\right)^{\nu}K_{\nu}(\sqrt{\mathbf{k}^{2}}\,\xi) \nonumber \\
&=\frac{\pi^{\frac{d}{2}+1}}{\Gamma\left(\Delta\right)\sin\left(\nu\pi\right)}\left(\left(\frac{1}{\xi^{2}}\right)^{\nu}\sum_{k=0}^{\infty}\frac{\left(\frac{\mathbf{k}^{2}\xi^{2}}{4}\right)^{k}}{k!\Gamma(-\nu+k+1)}-\left(\frac{\mathbf{k}^{2}}{4}\right)^{\nu}\sum_{k=0}^{\infty}\frac{\left(\frac{\mathbf{k}^{2}\xi^{2}}{4}\right)^{k}}{k!\Gamma(\nu+k+1)}\right).
\end{align}
If $\nu<0$, then the limit $\xi\rightarrow0$ of this result converges towards
\begin{equation}
G(\mathbf{k})=\lim_{\xi\rightarrow0}f_{\xi}(\mathbf{k})=-\frac{\pi^{\frac{d}{2}+1}}{\Gamma\left(\Delta\right)\Gamma(\nu+1)\sin(\nu\pi)}\left(\frac{\mathbf{k}^{2}}{4}\right)^{\nu}=\pi^{\frac{d}{2}}\frac{\Gamma\left(\frac{d}{2}-\Delta\right)}{\Gamma\left(\Delta\right)}\left(\frac{\mathbf{k}^{2}}{4}\right)^{\Delta-\frac{d}{2}}.
\end{equation}
This correspond exactly to the result in (\ref{ScalarFourierTransformDeltaLessThandOver2}). If $\nu>0$, the first terms will diverge in the limit $\xi\rightarrow0$. However, note that these terms are analytic in $\mathbf{k}^{2}$; they correspond to contact terms and are of no physical relevance. The only non-analytic term that survives in the limit $\xi\rightarrow0$ is given by
\begin{equation}
\tilde{G}(\mathbf{k})=-\frac{\pi^{\frac{d}{2}+1}}{\Gamma\left(\Delta\right)\Gamma(\nu+1)\sin(\nu\pi)}\left(\frac{\mathbf{k}^{2}}{4}\right)^{\nu}=\pi^{\frac{d}{2}}\frac{\Gamma\left(\frac{d}{2}-\Delta\right)}{\Gamma\left(\Delta\right)}\left(\frac{\mathbf{k}^{2}}{4}\right)^{\Delta-\frac{d}{2}}.
\end{equation}
This is the same result that we obtained by analytic continuation of (\ref{ScalarFourierTransformDeltaLessThandOver2}) in $\Delta$. If $n=\Delta-d/2$ is a positive integer, then we can use the series expansion
\begin{align}
K_{n}(x)&=\frac{1}{2}\left(\frac{x}{2}\right)^{-n}\sum_{k=0}^{n-1}\frac{(n-k-1)!}{k!}\left(-\frac{x^{2}}{4}\right)^{k}+(-1)^{n+1}\ln\left(\frac{x}{2}\right)I_{n}(x) \nonumber \\
&+\frac{1}{2}\left(-\frac{x}{2}\right)^{n}\sum_{k=0}^{\infty}\frac{\left(\psi(k+1)+\psi(n+k+1)\right)}{k!(n+k)!}\left(\frac{x^{2}}{4}\right)^{k}
\end{align}
to write (using also the series expansion of $I_{n}(x)$ as above),
\begin{align}
G_{\xi}(\mathbf{k})&=\frac{2\pi^{\frac{d}{2}}}{\Gamma\left(\Delta\right)}\left(\frac{\sqrt{\mathbf{k}^{2}}}{2\xi}\right)^{n}K_{n}(\sqrt{\mathbf{k}^{2}}\,\xi) \nonumber \\
&=\frac{\pi^{\frac{d}{2}}}{\Gamma\left(\Delta\right)}\left\{\left(\frac{1}{\xi^{2}}\right)^{n}\sum_{k=0}^{n-1}\frac{(n-k-1)!}{k!}\left(-\frac{\mathbf{k}^{2}\xi^{2}}{4}\right)^{k}-\ln\left(\frac{\mathbf{k}^{2}\xi^{2}}{4}\right)\left(-\frac{\mathbf{k}^{2}}{4}\right)^{n}\sum_{k=0}^{\infty}\frac{\left(\frac{\mathbf{k}^{2}\xi^{2}}{4}\right)^{k}}{k!\Gamma(n+k+1)}\right. \nonumber \\
&\qquad\qquad\left.+\left(-\frac{\mathbf{k}^{2}}{4}\right)^{n}\sum_{k=0}^{\infty}\frac{\left(\psi(k+1)+\psi(n+k+1)\right)}{k!(n+k)!}\left(\frac{\mathbf{k}^{2}\xi^{2}}{4}\right)^{k}\right\}.
\end{align}
In the limit $\xi\rightarrow0$, the first terms will diverge; however, these are analytic in $\mathbf{k}^{2}$, so we can neglect them. The last terms are also analytic in $\mathbf{k}^{2}$. The only non-analytic term that survives in the limit $\xi\rightarrow0$ is given by
\begin{equation}
\tilde{G}(\mathbf{k})=\frac{(-1)^{n+1}\pi^{\frac{d}{2}}}{\Gamma\left(\Delta\right)n!}\left(\frac{\mathbf{k}^{2}}{4}\right)^{n}\ln\left(\mathbf{k}^{2}\right)=\frac{(-1)^{\Delta-\frac{d}{2}+1}\pi^{\frac{d}{2}}}{\Gamma\left(\Delta\right)\Gamma\left(\Delta-\frac{d}{2}+1\right)}\left(\frac{\mathbf{k}^{2}}{4}\right)^{\Delta-\frac{d}{2}}\ln\left(\mathbf{k}^{2}\right).
\end{equation}
There is yet another way to find this last result. If $\Delta$ is such that $n=\Delta-d/2$ is an integer, we could also have used another regulator $\delta$ and replace everywhere $\Delta$ by $\Delta-\delta$. Then $n$ gets replaced by $n-\delta$, which is no longer integer. Now we can use that
\begin{align}
\Gamma\left(-n+\delta\right)&=\frac{(-1)^{n}}{n!}\frac{1}{\delta}\left(1+\psi(n+1)\delta+\mathcal{O}\left(\delta^{2}\right)\right), \\
\frac{1}{\Gamma\left(\Delta-\delta\right)}&=\frac{1}{\Gamma\left(\Delta\right)}\left(1+\psi(\Delta)\delta+\mathcal{O}\left(\delta^{2}\right),\right) \\
\left(\frac{\mathbf{k}^{2}}{4}\right)^{n-\delta}&=\left(\frac{\mathbf{k}^{2}}{4}\right)^{n}\left(1-\ln\left(\frac{\mathbf{k}^{2}}{4}\right)\delta+\mathcal{O}\left(\delta^{2}\right)\right),
\end{align}
such that
\begin{align}
\tilde{G}(\mathbf{k})&=\pi^{\frac{d}{2}}\frac{\Gamma\left(\frac{d}{2}-\Delta+\delta\right)}{\Gamma\left(\Delta-\delta\right)}\left(\frac{\mathbf{k}^{2}}{4}\right)^{\Delta-\frac{d}{2}-\delta} \nonumber \\
&=\frac{(-1)^{n}\pi^{\frac{d}{2}}}{\Gamma\left(\Delta\right)n!}\left(\frac{\mathbf{k}^{2}}{4}\right)^{n}\left(\frac{1}{\delta}+\psi(n+1)+\psi(\Delta)-\ln\left(\frac{\mathbf{k}^{2}}{4}\right)+\mathcal{O}\left(\delta\right)\right).
\end{align}
The term in $1/\delta$ is divergent and analytic. All terms of order $\mathcal{O}(\delta)$ vanish in the limit $\delta\rightarrow0$. The non-divergent term contains an analytic piece and a non-analytic one, namely
\begin{equation}
\tilde{G}(\mathbf{k})=\frac{(-1)^{n+1}\pi^{\frac{d}{2}}}{\Gamma\left(\Delta\right)n!}\left(\frac{\mathbf{k}^{2}}{4}\right)^{n}\ln\left(\mathbf{k}^{2}\right),
\end{equation}
which is exactly the right result.


\subsection{Minkowski signature}\label{integerminkowski}

We can perform a Wick rotation of the Euclidean two-point function
\begin{equation}
G(\mathbf{x})=\frac{1}{(\mathbf{x}^{2})^{\Delta}}=\frac{1}{(y^{2}+\vec{x}^{2})^{\Delta}},
\end{equation}
to Minkoswki signature by taking $y=e^{i\theta}t$ and continuously take $\theta$ from $0$ to $\pi/2-\epsilon$ (where $\epsilon>0$ and $\epsilon\rightarrow0$ is implied). We then find the time-ordered two-point function
\begin{equation}
iG_{F}(\mathbf{x})=\frac{1}{(-t^{2}+\vec{x}^{2}+i\epsilon)^{\Delta}}.
\end{equation}
For non-integer $\Delta$, we can rewrite this as
\begin{align}
iG_{F}(\mathbf{x})&=\frac{1}{(\vec{x}^{2}-t^{2})^{\Delta}}=\frac{\theta\left(\vec{x}^{2}-t^{2}\right)}{(\vec{x}^{2}-t^{2})^{\Delta}}+\frac{\theta\left(t^{2}-\vec{x}^{2}\right)}{(-(t^{2}-\vec{x}^{2}))^{\Delta}} \nonumber \\
&=\frac{\theta\left(\vec{x}^{2}-t^{2}\right)}{(\vec{x}^{2}-t^{2})^{\Delta}}+\frac{\theta\left(t^{2}-\vec{x}^{2}\right)}{(t^{2}-\vec{x}^{2})^{\Delta}}e^{-i\pi\Delta}.
\end{align}
For integer $\Delta$, one needs to pay attention to contact terms that may arise as for example in
\begin{align}
iG_{F}(\mathbf{x})=\frac{1}{\vec{x}^{2}-t^{2}+i\epsilon}=\mathcal{P}\frac{1}{\vec{x}^{2}-t^{2}}-i\pi\delta\left(\vec{x}^{2}-t^{2}\right).
\end{align}
In the mean text of this paper, we have assumed that $\Delta$ is non-integer. One can find the results of integer $\Delta$ by using the regularisation $\Delta-\delta$ as in the previous section, where the contact terms result in analytic terms in $\mathbf{k}^{2}$. One can show that this will change the expressions for the real part of the time-ordered and retarded propagators, but not the imaginairy part. Therefore, the expressions for the spectral function and occupation number are valid for all values of $\Delta$.

\setcounter{equation}{0}


\section{Complexified geodesics in AdS$_{3}$ and BTZ}\label{app:complex}

In this Appendix, we review the construction of complexified geodesics in AdS$_{3}$ and BTZ.


\paragraph{AdS$_{3}$}\mbox{}\\

In empty AdS$_{3}$ spacetime
\begin{equation}
ds^{2}=-r^{2}dt^{2} + \frac{dr^{2}}{r^{2}}+r^{2}dx^{2},
\end{equation}
the (real) geodesics are given by
\begin{itemize}
\item Spacelike geodesics
\begin{itemize}
\item[$\star$] $j^{2}>e^{2}$:
\begin{equation}
\begin{cases}
r(\lambda)=\sqrt{j^{2}-e^{2}}\cosh(\lambda-\lambda_{0}) \\
x(\lambda)=x_{0}+\frac{j}{j^{2}-e^{2}}\tanh(\lambda-\lambda_{0}) \\
t(\lambda)=t_{0}+\frac{e}{j^{2}-e^{2}}\tanh(\lambda-\lambda_{0})
\end{cases}
\end{equation}
\item[$\star$] $e^{2}>j^{2}$:
\begin{equation}
\begin{cases}
r(\lambda)=\sqrt{e^{2}-j^{2}}\sinh(\lambda-\lambda_{0}) \\
x(\lambda)=x_{0}-\frac{j}{e^{2}-j^{2}}\coth(\lambda-\lambda_{0}) \\
t(\lambda)=t_{0}-\frac{e}{e^{2}-j^{2}}\coth(\lambda-\lambda_{0})
\end{cases}
\end{equation}
\end{itemize}
\item Timelike geodesics
\begin{itemize}
\item[$\star$] $j^{2}>e^{2}$: $\qquad \qquad \qquad \qquad \qquad $no solutions
\item[$\star$] $e^{2}>j^{2}$:
\begin{equation}
\begin{cases}
r(\lambda)=\sqrt{e^{2}-j^{2}}\cos(\lambda) \\
x(\lambda)=x_{0}+\frac{j}{e^{2}-j^{2}}\tan(\lambda) \\
t(\lambda)=t_{0}+\frac{e}{e^{2}-j^{2}}\tan(\lambda)
\end{cases}\,.
\end{equation}
\end{itemize}
\end{itemize}
Timelike geodesics oscillate in the bulk and do not extend to the boundary. The first class of spacelike geodesics with $j^{2}>e^{2}$ go from the boundary into the bulk, and back to the boundary connecting spacelike separated points on the boundary.
The second class of spacelike geodesics with $e^{2}>j^{2}$ go from the boundary ($r\rightarrow\infty$) to the Poincar\'e horizon  ($r=0$). These are the ones that can be used to connect timelike separated points on the boundary. However, the expressions $x(\lambda)$ and $t(\lambda)$ diverge when the geodesic reaches the Poincar\'e horizon. To regularize this behavior, we add a small imaginary part to the affine parameter $\lambda$, so that the coordinates can thus in general take complex values:
\begin{equation}
\begin{cases} \label{eq:complexAdS}
r(\lambda)=\sqrt{e^{2}-j^{2}}\sinh(\lambda-\lambda_{0}+i\beta) \\
x(\lambda)=x_{0}-\frac{j}{e^{2}-j^{2}}\coth(\lambda-\lambda_{0}+i\beta) \\
t(\lambda)=t_{0}-\frac{e}{e^{2}-j^{2}}\coth(\lambda-\lambda_{0}+i\beta)
\end{cases}\,.
\end{equation}
These expressions can also be rewritten as
\begin{equation}
\begin{cases}
r(\lambda)=\sqrt{e^{2}-j^{2}}\left(\sinh(\lambda-\lambda_{0})\cos(\beta)+i\cosh(\lambda-\lambda_{0})\sin(\beta)\right) \\
x(\lambda)=x_{0}-\frac{j}{e^{2}-j^{2}}\left(\frac{\sinh(2(\lambda-\lambda_{0}))-i\sin(2\beta)}{\cosh(2(\lambda-\lambda_{0}))-\cos(2\beta)}\right) \\
t(\lambda)=t_{0}-\frac{e}{e^{2}-j^{2}}\left(\frac{\sinh(2(\lambda-\lambda_{0}))-i\sin(2\beta)}{\cosh(2(\lambda-\lambda_{0}))-\cos(2\beta)}\right)
\end{cases}\,,
\end{equation}
where we see that in order to have $x(\lambda)$ and $t(\lambda)$ finite everywhere, we need $\cos(2\beta)\neq1$.
Then:
\begin{eqnarray}
\Delta x& =&\lim_{\lambda\rightarrow\infty}x(\lambda)-\lim_{\lambda\rightarrow-\infty}x(\lambda)=-\frac{2j}{e^{2}-j^{2}}, \\
\Delta t& =&\lim_{\lambda\rightarrow\infty}t(\lambda)-\lim_{\lambda\rightarrow-\infty}t(\lambda)=-\frac{2e}{e^{2}-j^{2}},
\end{eqnarray}
such that $\Delta t^{2}-\Delta x^{2}=\frac{4}{e^{2}-j^{2}}$.

Note that for complexified $r$-coordinate, going from the boundary (at infinity) to the boundary (at infinity) corresponds to $|r(\lambda)|^{2}$ going from the point at infinity in the complex plane back to the point at infinity. The direction in which this happens is determined by the parameter $\beta$. We now rewrite the affine parameter $\lambda$ as a function of $|r|$:
\begin{eqnarray}
\lambda_{\pm}(|r|)&=&\lambda_{0}\pm\text{arccosh}\left(\sqrt{\frac{|r|^{2}}{e^{2}-j^{2}}+\cos^{2}(\beta)}\right)\,,
\end{eqnarray}
so that the geodesic length can be expressed as
\begin{equation}
\Delta\lambda(|r|)=\lambda_{+}(|r|)-\lambda_{-}(|r|)=2\ln\left(\sqrt{\frac{|r|^{2}}{e^{2}-j^{2}}+\cos^{2}(\beta)}+\sqrt{\frac{|r|^{2}}{e^{2}-j^{2}}-\sin^{2}(\beta)}\right).
\end{equation}
After subtracting the divergent part, we find the renormalised geodesic length
\begin{equation}
\delta\mathcal{L}_{\rm AdS} \equiv \lim_{|r|\rightarrow\infty}\left(\Delta\lambda(|r|)-2\ln(|r|)\right)=\ln\left(\frac{4}{e^{2}-j^{2}}\right)=\ln\left(\Delta t^{2}-\Delta x^{2}\right).
\end{equation}
This result is independent of the regularisation parameter $\beta$, and it agrees with our expectation for the scaling of two-point functions in a vacuum CFT.


\paragraph{BTZ}\mbox{}\\

We can repeat the construction in BTZ spacetime
\begin{equation}
ds^{2}=-(r^{2}-R^{2})dt^{2} +\frac{dr^{2}}{r^{2}-R^{2}}+r^{2}dx^{2}\,.
\end{equation}
For spacelike geodesics, we now distinguish the following cases:
\begin{itemize}
\item[$\star$] $(J^{2}-E^{2}+R^{2})^{2}>(2JR)^{2}$:
\begin{equation}
\begin{cases}
r^{2}(\lambda)=\frac{1}{2}\left((J^{2}-E^{2}+R^{2})+\cosh(2(\lambda-\lambda_{0}))\sqrt{(J^{2}-E^{2}+R^{2})^{2}-4J^{2}R^{2}}\right) \\
x(\lambda)=x_{0}+\frac{1}{R}\text{arccoth}\left(\left(\frac{J^{2}-E^{2}+R^{2}}{2JR}-\sqrt{\left(\frac{J^{2}-E^{2}+R^{2}}{2JR}\right)^{2}-1}\right)\tanh(\lambda-\lambda_{0})\right) \\
t(\lambda)=t_{0}+\frac{1}{R}\text{arccoth}\left(\left(\frac{J^{2}-E^{2}-R^{2}}{2ER}-\sqrt{\left(\frac{J^{2}-E^{2}-R^{2}}{2ER}\right)^{2}-1}\right)\tanh(\lambda-\lambda_{0})\right)
\end{cases}
\end{equation}
\item[$\star$] $(2JR)^{2}>(J^{2}-E^{2}+R^{2})^{2}$:
\begin{equation}
\begin{cases}
r^{2}(\lambda)=\frac{1}{2}\left((J^{2}-E^{2}+R^{2})+\sinh(2(\lambda-\lambda_{0}))\sqrt{4J^{2}R^{2}-(J^{2}-E^{2}+R^{2})^{2}}\right) \\
x(\lambda)=x_{0}+\frac{1}{R}\text{arccoth}\left(\frac{J^{2}-E^{2}+R^{2}}{2JR}\tanh(\lambda-\lambda_{0})-\sqrt{1-\left(\frac{J^{2}-E^{2}+R^{2}}{2JR}\right)^{2}}\right) \\
t(\lambda)=t_{0}+\frac{1}{R}\text{arccoth}\left(\frac{J^{2}-E^{2}-R^{2}}{2ER}\tanh(\lambda-\lambda_{0})-\sqrt{1-\left(\frac{J^{2}-E^{2}-R^{2}}{2ER}\right)^{2}}\right)
\end{cases}\,.
\end{equation}
\end{itemize}
Note that $\text{arccoth}(x)=\frac{1}{2}\ln\left(\frac{x+1}{x-1}\right)$ is only defined for $x\in\mathbb{R}\backslash]-1,1[$. Nevertheless, we can take the analytic continuation $\text{arccoth}(z)$ for $z\in\mathbb{C}\backslash\{-1,1\}$ which has a branch cut between the points 1 and -1.

To be able to take the AdS limit $R\rightarrow0$, we have to consider the first class of solutions with $(J^{2}-E^{2}+R^{2})^{2}>(2JR)^{2}$. In this first class, we can adopt the convenient notation
\begin{equation}
\Lambda_{\pm}=\frac{1}{2}\left((J^{2}-E^{2}+R^{2})\pm\sqrt{(J^{2}-E^{2}+R^{2})^{2}-4J^{2}R^{2}}\right),
\end{equation}
such that $\Lambda_{+}\Lambda_{-}=J^{2}R^{2}$ and $(\Lambda_{+}-R^{2})(\Lambda_{-}-R^{2})=E^{2}R^{2}$. We then find the complex solutions
\begin{equation}
\begin{cases} \label{eq:complexBTZ}
r^{2}(\lambda)=\Lambda_{-}+\left(\Lambda_{+}-\Lambda_{-}\right)\cosh^{2}(\lambda-\lambda_{0}+i\beta) \\
x(\lambda)=x_{0}+\frac{1}{R}\text{arccoth}\left(\sqrt{\frac{\Lambda_{-}}{\Lambda_{+}}}\tanh(\lambda-\lambda_{0}+i\beta)\right) \\
t(\lambda)=t_{0}+\frac{1}{R}\text{arccoth}\left(\sqrt{\frac{\Lambda_{-}-R^{2}}{\Lambda_{+}-R^{2}}}\tanh(\lambda-\lambda_{0}+i\beta)\right)
\end{cases}\,.
\end{equation}
The separation on the boundary is given by
\begin{equation}
\Delta x=\frac{2}{R}\text{arccoth}\left(\sqrt{\frac{\Lambda_{-}}{\Lambda_{+}}}\right)
\qquad\text{and}\qquad
\Delta t=\frac{2}{R}\text{arccoth}\left(\sqrt{\frac{\Lambda_{-}-R^{2}}{\Lambda_{+}-R^{2}}}\right),
\end{equation}
which gives
\begin{eqnarray}
\Lambda_{-}&=&\frac{R^{2}\cosh^{2}\left(\frac{R\Delta x}{2}\right)}{\sinh^{2}\left(\frac{R\Delta x}{2}\right)-\sinh^{2}\left(\frac{R\Delta t}{2}\right)} \,, \\
\Lambda_{+}&=&\frac{R^{2}\sinh^{2}\left(\frac{R\Delta x}{2}\right)}{\sinh^{2}\left(\frac{R\Delta x}{2}\right)-\sinh^{2}\left(\frac{R\Delta t}{2}\right)}\,.
\end{eqnarray}
We also have
\begin{equation}
\left|\frac{r^{2}-\Lambda_{-}}{\Lambda_{+}-\Lambda_{-}}\right|=\cosh^{2}(\lambda-\lambda_{0})-\sin^{2}(\beta)\,,
\end{equation}
which results into
\begin{equation}
\lambda_{\pm}=\lambda_{0}\pm\text{arccosh}\left(\sqrt{\left|\frac{r^{2}-\Lambda_{-}}{\Lambda_{+}-\Lambda_{-}}\right|+\sin^{2}(\beta)}\right)
\end{equation}
and
\bea
\delta\mathcal{L}_{\rm BTZ} &\equiv& \lim_{|r|\rightarrow\infty}\left(\Delta\lambda(|r|)-2\ln(|r|)\right) =\ln\left(\frac{4}{|\Lambda_{+}-\Lambda_{-}|}\right)\nonumber\\
&=&\ln\left[\frac{4}{R^2} \left(\sinh^{2}\left(\frac{R\Delta t}{2}\right)-\sinh^{2}\left(\frac{R\Delta x}{2}\right)\right)\right],
\eea
as expected.

\setcounter{equation}{0}


\section{A useful coordinate transformation of BTZ}\label{sec:coordinates}

The metric of a planar BTZ black hole is given by
\beq
ds^2=\frac{1}{z^2}\left[-(1-R^2z^2)dt^2+\frac{dz^2}{1-R^2z^2}+dx^2\right]. \label{eq:BTZ}
\eeq
It is locally identical to the metric of AdS$_3$ as can be seen by performing the change of variables
\begin{align}\label{eq:BTZAdS}
&\bar{x} \equiv \sqrt{R^{-2}-z^2}e^{Rx}\cosh(Rt)\,, \nonumber
\\
&\bar{t}\equiv \sqrt{R^{-2}-z^2}e^{Rx}\sinh(Rt)\,,
\\
&\bar z \equiv ze^{Rx}\,. \nonumber
\end{align}
The relations (\ref{eq:BTZAdS}) can be inverted to give
\begin{align}
&x=\frac{1}{2R}\log\left[ R^2(\bar z^2+\bar{x}^2-\bar{t}^2)\right]\,, \nonumber
\\
&t=\frac{1}{2R}\log\left(\frac{\bar{x}+\bar{t}}{\bar{x}-\bar{t}}\right)\,,
\\
&z=\frac{\bar{z}}{R\sqrt{\bar z^2+\bar{x}^2-\bar{t}^2}}\,, \nonumber
\end{align}
and the BTZ metric (\ref{eq:BTZ}) becomes
\beq
ds^2=\frac{1}{\bar z^2}(-d\bar{t}^2+ d\bar z^2+d\bar{x}^2).
\eeq

In particular, the retarded bulk-to-bulk propagator in the BTZ black hole is simply identical to the vacuum one in the new coordinates
\beq
iG_{BB,R}^{\rm BTZ}(\bar x,\bar t,  \bar z;  \bar x',\bar t',\bar{z}') = A \theta(\bar t-\bar t')\textrm{Im}\Big[\sigma^{-\Delta} \,_2F_1\Big(\frac{\Delta+1}{2},\frac{\Delta}{2};\Delta;\sigma^{-2}\Big)\Big],
\eeq
where $A$ is a real constant depending on $\Delta$ and we have defined
\beq
\sigma \equiv \frac{1}{2}\Big(\frac{\bar z'}{\bar z}+\frac{\bar z}{\bar z'}+\frac{(\bar x-\bar x')^2-(\bar t-\bar t')^2+i\epsilon}{\bar z \bar z'}\Big).\label{eq:btzbulkbulk}
\eeq
Using the definitions \eqref{eq:Gbdrybulk}-\eqref{eq:Gbulkbdry} given in Section~\ref{sect:joining},
we obtain the boundary-to-bulk propagator
\beq
iG_{B,R}^{\rm BTZ}(\ \bar x, \bar t, \bar z; \bar x',\bar t')=2^{\nu+2}\nu A\theta(\bar t-\bar t')\textrm{Im}\left\{\frac{\bar{z}^{\Delta}}{\left[-(\bar t-\bar t')^2+(\bar x-\bar x')^2+\bar z^2+i\epsilon\right]^{\Delta}}\right\},
\eeq
and the bulk-to-boundary propagator
\beq
i\tilde{G}_{B,R}^{\rm BTZ}(\bar x,\bar t;\bar x', \bar t', \bar z')=2^{\nu+2}\nu A\theta(\bar t-\bar t')\textrm{Im}\left\{\frac{(\bar{z}')^{\Delta}}{\left[-(\bar t-\bar t')^2+(\bar x-\bar x')^2+(\bar z')^2+i\epsilon\right]^{\Delta}}\right\}.
\eeq

\setcounter{equation}{0}


\section{Relating $G_R$ to the one-point function}\label{sec:onepoint}

Consider a quantum system described by the following Hamiltonian
\beq
H=H_0-J(t)\mathcal{O}\,,
\eeq
where we have suppressed the spatial coordinates as they play no crucial role in the following analysis.
For concreteness we work in the Schrodinger representation where the only operator depending on time
is the density matrix. In the following we will take $H_0$ to be a generic Hamiltonian which is allowed
to have explicit time dependence.

Let us assume we have solved for the density matrix $\rho$ when $J=0$, meaning that we know the general
solution to the equation
\beq
i\partial_t\rho_0(t)=\[H_0,\rho_0(t)\].
\eeq
The solution $\rho_0(t)$ can be written in terms of a time evolution operator $U$, which is defined by
\beq
i\partial_t U(t)=H_0 U(t),
\eeq
so that
\beq
\rho_0(t)=U(t)\rho_0(t=0)U^{\dagger}(t).
\eeq
The idea of linear response is to work out the time dependence of the density operator to linear order in $J(t)$.
Consider the ansatz $\rho=\rho_0+\delta\rho$, where we assume $\delta\rho$ to be of the order of $J$. The equation
of motion for $\delta\rho$ at linear order becomes
\beq
i\partial_{t}\delta\rho(t)=\[H_0,\delta\rho(t)\]-J(t)\[\mathcal{O},\rho_0(t)\].
\eeq
The solution can be obtained through the method of variation of constant $\delta\rho=U s U^{\dagger}$, which allows to integrate
the equation
\beq
s(t)=s(t_0)+i\int_{t_0}^{t}dt'J(t')\[\mathcal{O}_H(t'),\rho_0(t')\],
\eeq
where we have defined the Heisenberg picture operator $\mathcal{O}_H(t)=U(t)\mathcal{O}U^{\dagger}(t)$. Assuming $J(t)$ vanishes for $t\leq t_0$ for some $t_0$ we get the result for the density matrix
\beq
\rho(t)=\rho_0(t)+i\int_{-\infty}^{t}dt'J(t')U(t)\[\mathcal{O}_H(t'),\rho_0(t')\]U^{\dagger}(t).
\eeq
The expectation value of the operator $\mathcal{O}$ becomes
\beq
\langle\mathcal{O}(t)\rangle_J=\textrm{Tr}\Big(\rho(t)\mathcal{O}\Big)=\textrm{Tr}\Big(\rho_0(t)\mathcal{O}\Big)
+i\int_{-\infty}^{t}dt'J(t')\textrm{Tr}\Big(\rho_0(t)\[\mathcal{O}_H(t),\mathcal{O}_H(t')\]\Big),
\eeq
so that the overall effect of the source $J$ on the expectation value of $\mathcal{O}$ is
\beq
\langle\mathcal{O}(t)\rangle_J-\langle\mathcal{O}(t)\rangle_0=-\int_{-\infty}^{\infty}dt'J(t')G_R(t,t').
\eeq
For the specific source $J(t)=\epsilon\delta(t-t_0)$, one obtains
\beq
\langle\mathcal{O}(t)\rangle_{\epsilon \delta}-\langle\mathcal{O}(t)\rangle_0=-\epsilon G_R(t,t_0).
\eeq
Dividing by $\epsilon$ and taking the limit $\epsilon\rightarrow 0$ gives
\beq
\frac{\partial}{\partial\epsilon}\langle\mathcal{O}(t)\rangle_{\epsilon  \delta}=-G_R(t,t_0).
\eeq
On the other hand, in the holographic setup the one-point function in the presence of a source of the form $J=\epsilon\delta(t-t_0)\delta(x-x_0)$
is given by
\beq
\langle\mathcal{O}(x,t)\rangle_{\epsilon \delta}=\frac{\delta S^{\rm on-shell}}{\delta J(x,t)}=-2\nu\phi_+ \,,
\eeq
where $\phi= \epsilon \delta(t-t_0)\delta(x-x_0) z^{\Delta_-}+z^{\Delta_+}\phi_+ + ...$ as $z \to 0$. The dependence of $\phi_+$ on $\epsilon$
is linear since the bulk equation of motion is linear (in the leading large-$N$ approximation). Thus, $\phi_+/\epsilon$
is independent of $\epsilon$ and we have
\beq
\frac{\partial}{\partial\epsilon}\langle\mathcal{O}(x,t)\rangle_{\epsilon \delta}=2\nu\frac{\phi_+}{\epsilon}.
\eeq
Since the final result is independent of $\epsilon$ we can set $\epsilon=1$, which gives
\beq
-G_R(x,t;x_0,t_0) = \left. \frac{\partial}{\partial\epsilon}\langle\mathcal{O}(x,t)\rangle_{\epsilon \delta} \right|_{\epsilon =1}=\langle\mathcal{O}(x,t)\rangle_{\delta}\,,
\eeq
where we have reinstated the spatial dependence. 

Notice that the assumption that $\phi_+$ is a linear function of $\epsilon$ can fail if the operator $\mathcal{O}$ has a non-vanishing expectation value. We will not consider such a 
case further in this paper.

\setcounter{equation}{0}


\section{AdS$_3$ retarded bulk-to-boundary propagator}\label{sec:retarded}

The Euclidean AdS boundary-to-bulk propagator is given by
\beq
G_{E,B}^{\rm AdS}(x,y,z;x',y')=C_{\Delta}\frac{z^{\Delta}}{\left[z^2+(y-y')^2+(x-x')^2\right]^{\Delta}}\,,
\eeq
where
\beq
C_{\Delta}=\frac{\Gamma(\Delta)}{\pi\Gamma(\Delta-1)}\,.
\eeq
This can be Fourier transformed into a mixed momentum space as
\beq
G_{E,B}^{\rm AdS}(k,y,z;y')=C_{\Delta}\int_{-\infty}^\infty dx\frac{z^{\Delta}e^{-ikx}}{(z^2+(y- y')^2+x^2)^{\Delta}}\,,
\eeq
One way to perform this integral is to use the Schwinger parametrization
\beq
G_{E,B}^{\rm AdS}(k,y,z;y')=\frac{C_{\Delta}}{\Gamma(\Delta)}\int_{-\infty}^\infty dx \int_0^{\infty}ds\, s^{\Delta-1}z^{\Delta}e^{ikx-s(x^2+(y-y')^2+z^2)}\,,
\eeq
and perform the Gaussian integral over $x$ to get
\beq
G_{E,B}^{\rm AdS}(k,y,z;y')=\frac{C_{\Delta}z^{\Delta}}{\Gamma(\nu+1)}\int_0^\infty ds\, s^{\nu-\frac 1 2}e^{-s a-\frac{k^2}{4s}}\,,
\eeq
where we have introduced $a=z^2+(y-y')^2$. Introducing a new integration variable $u=as$ we get
\beq
G_{E,B}^{\rm AdS}(k,y,z;y')=\frac{C_{\Delta}z^{\Delta}\sqrt{\pi}}{\Gamma(\nu+1)}a^{-\nu-\frac 1 2}\int_0^\infty du\, u^{\nu-\frac 1 2}e^{-u-\frac{k^2a}{4u}}\,,
\eeq
which can be recognized as an integral representation of a Bessel function
\beq \label{eq:boh}
G_{E,B}^{\rm AdS}(k,y,z;y')=\frac{C_{\Delta}\sqrt{\pi}2^{\frac{1}{2}-\nu}}{\Gamma(\nu+1)}
z^{\nu+1}|k|^{\nu+\frac{1}{2}}a^{-\frac{1}{2}(\nu-\frac{1}{2})}K_{-\nu- \frac 1 2}(|k|\sqrt{a})\,.
\eeq
Next we want to compute the retarded propagator, which can be obtained from the Feynman one through the identity
\begin{align}
G_{B,R}(k,t,z;t')=-\theta(t-t')\Big( G_{B,F}(k,t,z;t')+ G_{B,F}^*(k,t,z;t')\Big)\,.
\end{align}
To determine the Feynman propagator in the future lightcone, we should continue \eqref{eq:boh} to real time as $a=z^2+(y-y')^2
\rightarrow -(t-t')^2+z^2+i\epsilon$. Since we need fractional powers of $a$ we need to take into account
the $i\epsilon$ factor to determine the phase $a^{\alpha}\rightarrow (-(t-t')^2+z^2+i\epsilon)^{\alpha}=e^{i\pi\alpha}b$, where $b$ is the
positive quantity $b=(t-t')^2-z^2$. 
In continuing to real time the Feynman propagator is multiplied by a factor of $i$ as a part of its definition. In this
way we obtain 
\beq
G_{B,F}^{\rm AdS}(k,t,z;t')= \frac{C}{2}z^{1+\nu}|k|^{\nu+\frac 1 2}b^{-\frac{1}{2}(\nu+\frac{1}{2})}H^{(2)}_{-\nu-\frac{1}{2}}(|k|\sqrt{b})\,,
\eeq
where we have defined a new normalization constant
\beq
C=\frac{2^{\frac{1}{2}-\nu}\sqrt{\pi}}{\Gamma(\nu)}\,.\label{eq:constant}
\eeq
The retarded bulk-to-boundary propagator can be obtained from the Feynman propagator through the identity
\begin{align}
&G_{B,R}^{\rm AdS}(k,t,z;t')=- C\theta(t-t'-z)z^{1+\nu}|k|^{\nu+\frac 1 2}b^{-\frac{2\nu +1}{4}}J_{-\nu-\frac{1}{2}}(\sqrt{b}|k|),
\end{align}
where the $\theta(t-t'-z)$ appears because $ G_{B,F}^{\rm AdS}$ is purely imaginary for $b<0$, which is outside the future lightcone.

\setcounter{equation}{0}



\end{document}